\documentclass[12pt]{report}
\input{urreport.sty}
\usepackage[english]{babel}
\usepackage{alltt}
\usepackage[pdftex]{graphicx}
\usepackage[pdftex]{epsfig}
\usepackage{graphics}
\usepackage{amsthm}
\usepackage{amsmath}
\usepackage{amssymb}
\usepackage{algorithm}
\usepackage{algorithmic}
\usepackage{fancyvrb}

\usepackage{hyperref}

\usepackage{anyfontsize}

\usepackage{caption}
\newcommand{\cbstart}{
	\begin{flushright}
	\begin{tabular}{|l|l|}
	\hline
	\makebox[48pt][l]{\em type}&\makebox[330pt][l]{\em command}\\
	\hline
}
\newcommand{\cbfin}{
	\end{tabular}
	\end{flushright}
}

\newcommand{\bp}{\begin{picture}(0,0)}

\newcommand{\fp}{\end{picture}}

\newcommand{\citep}[1] {\cite{#1}}

\begin{document}
\setlength{\parskip}{0pt}
%
%
%
%
%


\title{\fontsize{45}{33}\selectfont Distributed Computing From First Principles}

\author{\textbf{Kenneth Emeka Odoh}}
\date{August 2024}

\beforepreface
\chapter*{Preface}

\subsubsection{"I would maintain that thanks are the highest form of thought; and that gratitude is happiness doubled by wonder." -- G.K. Chesterton}
\subsubsection{"An adventure is only an inconvenience rightly considered. An inconvenience is only an adventure wrongly considered." -- G.K. Chesterton}

This work is \href{https://kenluck2001.github.io/blog_post/authoring_a_new_book_on_distributed_computing.html}{motivated} by my quest to acquire comprehensive knowledge in a technical field with practical real-world applications. Our manuscript aims to inspire the next generation of software practitioners (Engineers, scientists, and physicists) to apply distributed computing paradigms to address their challenges.

I am grateful to the numerous reading groups in Vancouver that spurred my interest in Distributed Systems. Despite my humble beginnings, I am now privileged to have developed into a seasoned Software Engineer. This book represents my opportunity to contribute back to society. Writing this book has been the most challenging endeavor of my evolving career. Hence, this work is a \textbf{charitable undertaking} and will be \textbf{royalty-free} to maximize the benefit for underrepresented minorities.

While living mentors are invaluable in my career development, these late individuals had the greatest influence on my life: Thomas Aquinas (bright star of the \href{https://en.wikipedia.org/wiki/Dominican_Order}{Dominican Order}), Chinua Achebe (chronicler of \href{https://en.wikipedia.org/wiki/Igbo_people}{Igbo} civilization), G.K. Chesterton (prince of paradox), and Tupac Amaru Shakur (thug poetry). In the words of G.K. Chesterton, "A good novel tells us the truth about its hero; but a bad novel tells us the truth about its author." Therefore, the judgment rests with the readers~\footnote{The motivation for this book is detailed at \url{https://kenluck2001.github.io/blog_post/authoring_a_new_book_on_distributed_computing.html}.}~\footnote{The front page illustration~\ref{boobktitle}was created by Rofiat Ibrahim using pencil on paper. Rofiat is a talented artist and a student whom I met by chance at Yaba College of Technology in Lagos, Nigeria, in December 2022. Her excellent drawing features a broken kola nut and a set of palm wine jars, items of profound cultural significance in \href{https://en.wikipedia.org/wiki/Igbo_people}{Igbo} tradition. This figure crystallizes my thought that knowledge should be "\textbf{free as in free beer}".}.


\chapter*{Acknowledgement}

\subsubsection{ \small "When it comes to life the critical thing is whether you take things for granted or take them with gratitude." -- G.K. Chesterton}

Archit Goyal reached out and provided source code on CRDT among other major reviewing and coding contributions. Archit is a Senior Software Engineer at LinkedIn, deeply involved in building and productionizing large-scale distributed data systems. He is one of the early engineers behind LinkedIn’s Flink Batch platform, powering Ads AI workloads and high-throughput, low-latency pipelines. Much of his work focuses on scaling stateful computations, disaggregated shuffle, checkpointing trade-offs, failure recovery, resource fairness, and multi-cluster observability.

I want to thank Anselm Eickhoff and Charles Krempeaux, who were responsive to my questions about CRDT. Also, I would also like to thank Harrington Joseph, who suggested that I focus on gaining a low-level understanding of parallel programming. We provided credits for both anonymous reviewers that provided tangible support in making this \href{https://kenluck2001.github.io/blog_post/book_launch_published_a_free_technical_book_on_distributed_computing.html}{manuscript}.

Kindly make a free-will donation to support my work, including open-source development, blogging, research papers, or textbooks. Thank you so much for your financial support.

\textbf{Donate}:     \url{https://buymeacoffee.com/kenluck2001}

\afterpreface
\setlength{\parskip}{0pt}
\chapter{Introduction}
\label{Introduction}
\subsubsection{ \small "What we know is a drop, what we don't know is an ocean." -- Isaac Newton}

Have you ever wondered how a typical distributed system works under the hood? Are you looking for a pedagogical guide with complete implementations and tricks of the trade? Look no further and read my writing on the topic. We have implemented several foundational algorithms in Distributed Computing. This paradigm has become ubiquitous in the industry where multiple systems interact to solve computational tasks, with applications including distributed stores (databases), IoT sensor networks, and the Internet. The exponential growth in deploying Distributed systems for solving real-world problems has led to a resurgence in understanding Distributed Computing from scratch~\footnote{ \footnotesize My research in Distributed Systems is a two-phased process: The first phase was covered in the \url{https://kenluck2001.github.io/blog_post/distributed_computing_from_first_principles.html}. The second phase is the current book that you are reading. A multi-phased approach to research has helped to minimize risk.}. Consensus is a foundational requirement for distributed systems, allowing disparate nodes to synthesize a consistent global perspective from their local operations. This collaborative agreement forms the very objective of distributed algorithms.

Distributed systems operate in inherently uncertain and chaotic environments, facing network unreliability through intermittent connectivity, message loss, and out-of-order delivery. Therefore, fault tolerance must be a fundamental architectural requirement, not an afterthought. As a result, managing these complexities demands rigorous correctness guarantees to prevent unintended behaviors at scale across multiple nodes and unreliable channels. In production-grade software, formal verification is crucial to reduce bugs and guarantee safety and liveness properties are never violated, especially in mission-critical applications.

A few years ago, my deficiency in the low-level knowledge required to build a large-scale distributed system without utilizing third-party packages motivated me to research this topic from first principles. I subsequently became a better Software Engineer in the process. My initial foray into distributed systems began by participating in a Distributed System meetup in Vancouver, where I eventually became the organizer. Our study group utilized materials from the Distributed Systems course at \href{https://www.kth.se/en}{KTH} Sweden. Our goal was mainly to gain intuition and theoretical knowledge on the subject. Fortunately, I took a further step by completing the Parallel, Concurrent, and Distributed Programming in Java Specialization on Coursera, where OpenMPI was mentioned. I recognized this as the missing link in my exploration of implementing low-level distributed algorithms. 

As I started researching, I noticed a troubling trend in Distributed Systems research: the most significant work in the field tends to be done in top research labs, advanced technical schools, and by seasoned open-source contributors. Unfortunately, this status quo is unacceptable. Hence, I am motivated to invest in this research to distill this knowledge to a diverse audience. Our focus is on delivering scientific content without diluting its quality and maintaining world-class rigor. We aim to create disruptive change by sharing knowledge hidden in plain sight. One axiom from the Zen of Python posits that "practicality beats purity," so rather than pontificate on the state-of-the-art distributed algorithms, we have adopted the approach of solidifying the fundamentals.

~\textbf{Notes}:
\begin{enumerate}

\item This work is based on non-proprietary content unrelated to my employer. The \href{https://kenluck2001.github.io/blog_post/distributed_computing_from_first_principles.html}{blog} summarizes the first phase of my two-phased research on low-level Distributed systems. The second phase is completed in this textbook.

\item All source code is the original work of the author.

\item Unlike other intellectual works, some of our references will be secondary sources. This decision is taken to keep our references within a reasonable scope.
\end{enumerate}

Full disclosure: I implemented portions of the KTH Distributed system course that ran on Edx, taught by Professor Haridi. I am grateful as it was my first exposure to Distributed computing. I will also include some of his slides (Paxos, sequence Paxos) in this book with full attribution.

~\textbf{Terminologies}

\begin{enumerate}
\item Quorum: a set containing a majority of correct processes.
\item Nodes, replicas, and processes will be used interchangeably and have the same meaning within the context of this book.
\item Model: an abstraction of the dynamics of a system.
\item Reliable broadcast: a message sent to a group of processes is delivered to all or none.
\item Atomic commit: processes either all commit or all abort a transaction.
\item Transaction: a series of operations that must be completed without interruption. If any operation fails, the entire transaction is aborted, e.g., 2-phase commit, 3-phase commit.
\item Multicast: a single source sends to multiple destinations and operates in LAN and WAN environments.
\item Broadcast: a single source sends to multiple destinations and operates in a LAN environment.
\item Computational graph: an abstraction of the ordered sequence of processes.
\item Optimistic concurrency is a suitable strategy when low contention is expected. Computation on a shared object is expensive compared to the overhead of locks in such scenarios.
\item RAID: This storage paradigm, known as a Redundant Array of Independent Disks, provides fault tolerance in the face of disk failures.
\item Volume: This refers to the available storage space accessible by the operating system.
\end{enumerate}

\section{Algorithm Implementations }
\label{Algorithm-Implementations}

My source code was tested with the following specifications:

\begin{tabular}{ |p{6cm}||p{6cm}| }
 \hline
Setup & Version\\
 \hline
Operating system    & Ubuntu 16.04.6 LTS \\
GCC                 & 5.4.0 20160609 \\
OpenMPI             & 1.10.2 \\
Python             & 2.7 \\
 \hline
\end{tabular}

Here is my list of implemented Distributed algorithms with associated source codes. They include:

\begin{tabular}{ |p{0.6cm}||p{6cm}|p{6cm}|  }
 \hline
 S/N & Algorithms & Source codes\\
 \hline
 1 & Logical clock    & \href{https://github.com/kenluck2001/DistributedSystemReseach/blob/master/textbook/vector2.c}{vector2.c}, \href{https://github.com/kenluck2001/DistributedSystemReseach/blob/master/textbook/lamport1.c}{lamport1.c} \\
 2 & Paxos (with Snapshot)  & \href{https://github.com/kenluck2001/DistributedSystemReseach/blob/master/textbook/single-paxos3-snapshot.c}{single-paxos3-snapshot.c}\\
 3 & Sequence Paxos   & \href{https://github.com/kenluck2001/DistributedSystemReseach/blob/master/textbook/sequence-paxos4.c}{sequence-paxos4.c} \\
 4 & Failure detector & \href{https://github.com/kenluck2001/DistributedSystemReseach/blob/master/textbook/failure-detector.c}{failure-detector.c} \\
 5 & Leader election  & \href{https://github.com/kenluck2001/DistributedSystemReseach/blob/master/textbook/leader-election3.c}{leader-election3.c}\\
 6 & Grow-only Counter CRDT  & \href{https://github.com/kenluck2001/DistributedSystemReseach/blob/master/textbook/crdt-gcounter.c}{crdt-gcounter.c}\\
 7 & Last-Write-Wins Map CRDT  & \href{https://github.com/kenluck2001/DistributedSystemReseach/blob/master/textbook/lww-map.c}{lww-map.c}\\
 8 & Raft             & Raft algorithm in Subsection~\ref{raft} \\
 9 & Distributed shared primitives  & shared primitive in Subsection~\ref{distributed-computing-patterns}\\
 10 & Distributed hashmap            & \href{https://github.com/kenluck2001/DistributedSystemReseach/blob/master/textbook/lamport1-majority-voting8.c}{lamport1-majority-voting8.c}\\
 11 & Two-Phase Commit          & \href{https://github.com/kenluck2001/DistributedSystemReseach/blob/master/textbook/two-phase-commit.c}{two-phase-commit.c}\\
 12 & Autoscaling Framework  (Serveless)             & \href{https://github.com/kenluck2001/DistributedSystemReseach/blob/master/textbook/autoscaling.py}{autoscaling.py} in Subsection~\ref{autoscaling-casestudy}\\

 \hline
\end{tabular}

\textbf{Source Code}: \href{https://github.com/kenluck2001/DistributedSystemReseach/tree/master/textbook}{click here}

Some of my implementations are in the playground folder, where you can follow my thought process during incremental development. Some of the implementations in the playground contain faulty solutions due to incorrect assumptions. However, the implementations referenced directly in this book are verified by the author and relatively free of obvious errors.
We settled on OpenMPI as the underlying library to provide low-level functionality. This choice is preferable to using a Remote Procedural Call, which obscures the message-passing paradigm by passing arguments as messages. Although conceptually similar, MPI provides a structured way of implementing our Distributed algorithms. Hence, all implementations will strictly use OpenMPI in C. Event-based programming based on message passing serves as a powerful abstraction for building Distributed systems. Therefore, we are using this approach for all of our implementations.
For the sake of pedagogy, we have omitted tracing and invariant proofs. While these are important in Distributed systems scholarship, we anticipate that users will learn by doing and not be hindered by excessive mathematics. However, for those who wish to focus more on writing proofs, we recommend this resource~\cite{lynch1993}.

\section{Overview }
\label{overview}

This book is organized into a number of Chapters. We begin by providing motivation, refreshers on prerequisites, foundational theory and algorithms, formal verification, and exercises.

Chapter~\ref{Introduction} motivates embarking on the adventure of understanding the underpinnings of Distributed computing from scratch. This section includes a list of implemented algorithms and sets the stage for the entire book.

Chapter~\ref{Preq-Dist-Comp} provides a quick primer on parallel programming and network programming concepts that are fundamental for forming the correct mental model required to understand more advanced concepts in the book. It is in this section that we introduce OpenMPI as the message-passing library to support our implementation of Distributed algorithms.

Chapter~\ref{Basic-Dist-Comp} describes Distributed systems. It also discusses the CAP theorem, the FLP impossibility of consensus, the Two Generals' Problem, accounting for the absence of a global clock through the introduction of logical clocks, and a principled way to detect failures.

Chapter~\ref{Dist-Con-Algo} describes foundational Distributed algorithms such as Paxos, Raft, election algorithms, and stabilization algorithms.

Chapter~\ref{Anti-Entropy} describes anti-entropy techniques. This is critical for achieving a consistent state across nodes, as divergences can affect agreement in these systems. The approach focuses on determining divergence across replicas, a fault-tolerant way of propagating changes to enforce a consistent state across replicas, reducing the time needed for eventual consistency, and performing error correction.

Chapter~\ref{Peer-to-Peer-Conn} describes peer-to-peer networks. This helps to create fault-tolerant systems that do not depend on a single server, as central servers are a source of single points of failure. Building peer-to-peer networks can help produce the redundancy needed to achieve a system where a single server is not a bottleneck.

Chapter~\ref{Formal-Verification} describes the formal verification of Distributed systems. It is imperative that in mission-critical systems, safety and liveness properties should not be violated at any time.

Chapter~\ref{Miscellaneous} describes topics such as distributed commit, coding philosophy, case studies, practical considerations for deploying distributed systems in the wild, testing strategies, evaluation metrics, and exercises.

\chapter{Prequisites For Distributed Computing}
\label{Preq-Dist-Comp}
\subsubsection{ \small "He who knows all the answers has not been asked all the questions." -- Confucius } 
\subsubsection{ \small "Anything worth doing is worth doing badly" -- G.K. Chesterton}

We have demonstrated utilizing the message-passing paradigm for implementing Distributed Systems in this work. The requirement for nodes to exchange information over a channel, warrants a solid understanding of network programming to grasp packet dynamics as information flows across the channel. Networking programming skills are closely complemented with the knowledge of Parallel programming. As a result of these prerequisites, we provide a brief introduction to both subjects in this manuscript. This knowledge is sufficient to understand the internals of real-world Distributed systems, as covered in Sections \ref{Network-Programming} and \ref{Parallel-Programming}.

\section{Network Programming}
\label{Network-Programming}
The socket is the fundamental file descriptor for networking. Recalling the Unix philosophy that everything is a file, the `socket()` function returns a descriptor. Sending a message over a network becomes analogous to writing to a file stream, while receiving a message is comparable to reading a file stream. However, in this context, the network acts as the channel for this communication.
There are two basic socket types, as outlined by~\cite{brian2020}:

\begin{itemize}
\item Stream socket (`SOCK\_STREAM`)
\item Datagram socket (`SOCK\_DGRAM`)
\end{itemize}

Most local networks utilize internal IP addresses, whereas Internet-facing gateways employ external IP addresses with Network Address Translation (NAT) performing the conversion from the Internet IP to the internal IP.

A server typically operates in a listening mode. Clients send messages to the server to trigger actions, resulting in a response to the caller. Socket programming is a  method for machines to connect over a network. This communication occurs via a protocol using a set of predefined conventions that enable parties to understand message structures, codebooks, and other auxiliary information necessary for the synthesis of exchanged data.

\subsubsection{Network Protocol}
Network protocols support varying degrees of error recovery, fault tolerance, and overhead rates for message transfer. The design requirement should inform the choice of the network protocol.Be it Transmission Control Protocol (TCP)~\footnote{\url{https://datatracker.ietf.org/doc/html/rfc9293}}, User Datagram Protocol (UDP)~\footnote{\url{https://datatracker.ietf.org/doc/html/rfc768}}, Quick UDP Internet Connections (QUIC)~\footnote{\url{https://datatracker.ietf.org/doc/html/rfc9002}}, or Licklider Transmission Protocol (LTP)~\footnote{\url{https://datatracker.ietf.org/doc/html/rfc5326}} respectively.

\begin{figure}[H]
\centering
\includegraphics[scale=0.4]{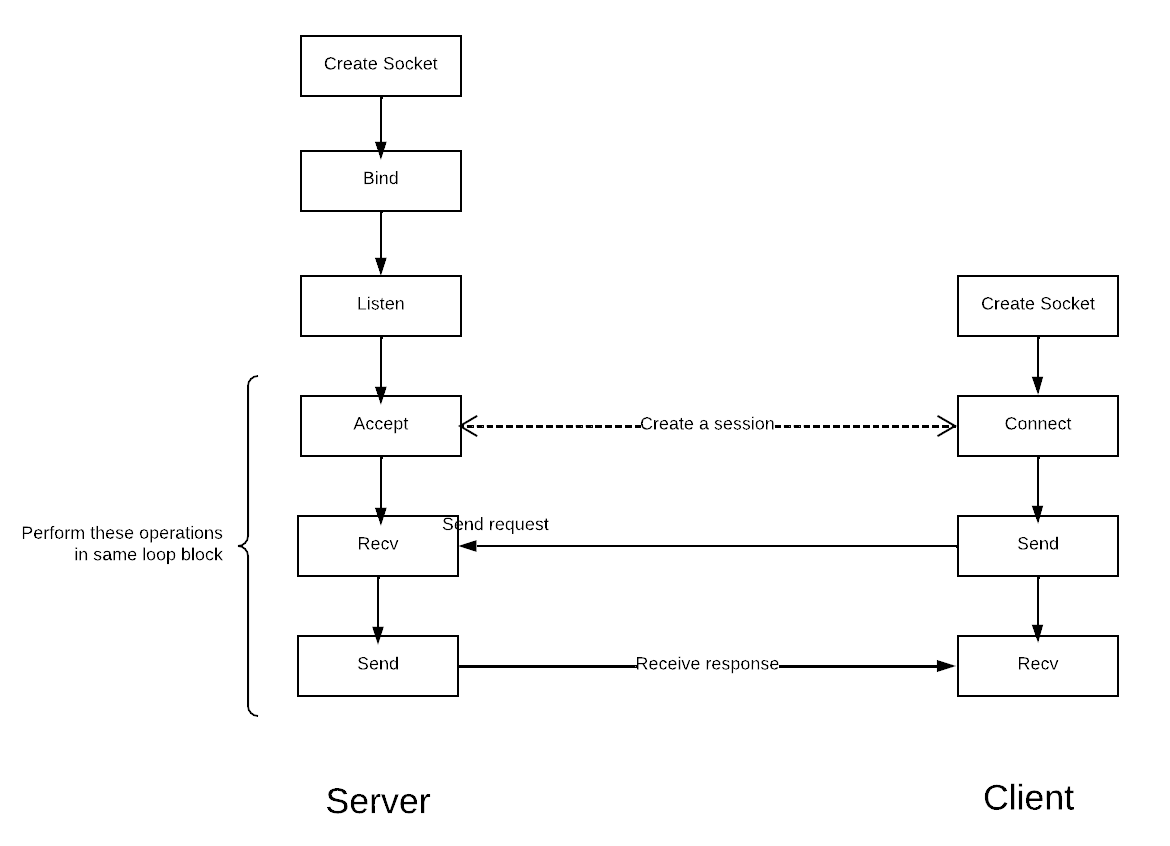}
\caption{ Sequence diagram for TCP.}
\label{tcp-flow-diag}
\end{figure}

Transmission Control Protocol (TCP) is a connection-oriented protocol operating at the transport layer. This protocol requires a dedicated connection must be established between the sender and receiver before data can be exchanged. A key characteristic of TCP is its provision of congestion control mechanisms, which dynamically adjust the transmission rate to avoid network overload. Furthermore, TCP guarantees reliable and order-preserving delivery of packets, ensuring that all data arrives at the destination correctly and in the order in which packets were sent. However, these features contribute to TCP being considered a "heavyweight" protocol due to the overhead associated with connection management and reliability mechanisms. Error recovery gets handled by the protocol, where lost messages are re-sent. The TCP protocol described in Figure~\ref{tcp-flow-diag}.

\begin{figure}[H]
\centering
\includegraphics[scale=0.4]{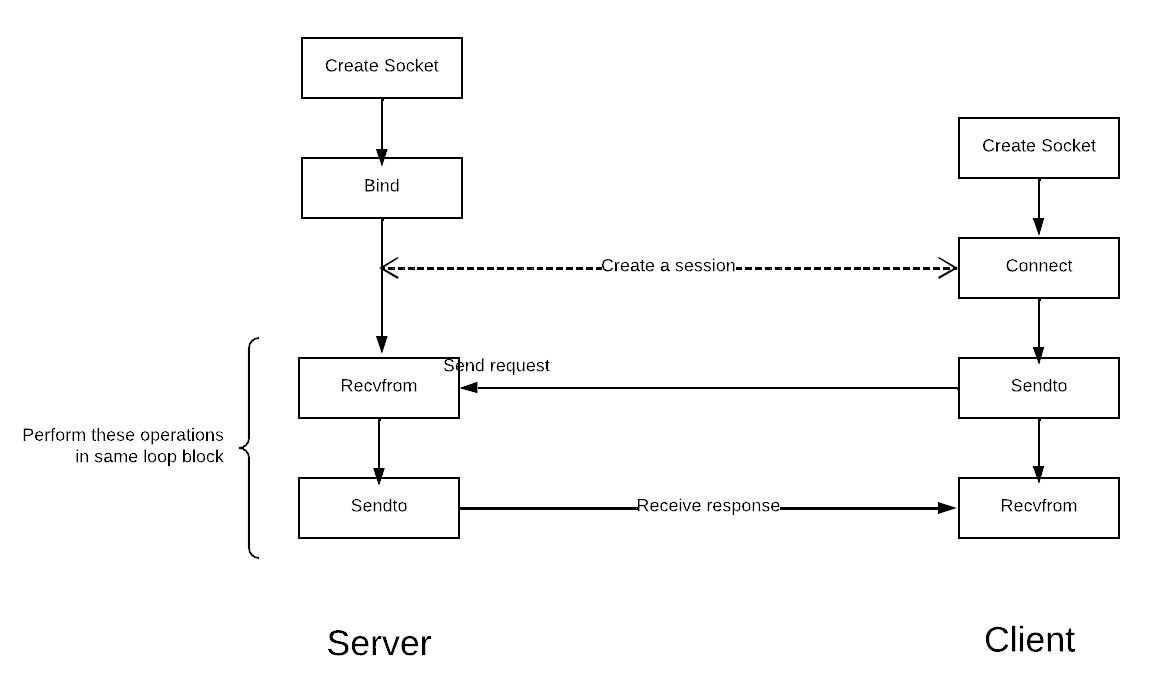}
\caption{ Sequence diagram for UDP.}
\label{udp-flow-diag}
\end{figure}

User Datagram Protocol (UDP) is a connectionless-oriented protocol, that operates at the transport layer. Unlike TCP, UDP does not require a established connection before communication begins. Consequently, it is lacking the built-in congestion control and reliability guarantees of TCP, meaning packets may be lost, duplicated, or arrive out of order. The absence of these features makes UDP a "lightweight" protocol with lower overhead, making it suitable for applications where speed and low latency gets prioritized over guaranteed delivery. Error recovery is handled by secondary protocols implemented on top of UDP to provide resilience. The protocol gets described in Figure~\ref{udp-flow-diag}.

Quick UDP Internet Connections (QUIC) represents an evolution of UDP. While built upon UDP, QUIC incorporates mechanisms to provide reliability and security features associated with TCP, such as reliable, in-order delivery and encryption. A significant advantage of QUIC is its ability to achieve these guarantees with reduced connection establishment time and lower overhead compared to TCP, making it a more efficient option for many applications.

Licklider Transmission Protocol (LTP) is a lightweight protocol that operates at the data link layer, a lower layer in the network stack compared to TCP and UDP. LTP is designed for challenging network environments characterized by intermittent connectivity and long delays, such as deep space communication. It does not implement congestion control and allows for opportunistic message transfer, where sending might be delayed until a suitable communication channel becomes available. LTP is a hybrid protocol that can be configured to provide  reliable or unreliable delivery depending on the requirements of the application.

The application requirement should guide the choice of the appropriate network protocol in the design of a Distributed System. For more information, refer to the illustration in this \href{https://workingwithruby.com/wwtcps/evented/}{blog}.

When communicating between nodes in Distributed systems, a channel is necessary to provide a medium for message transfer. The types of channels in use in Distributed systems~\cite{petrov2019} include:

\begin{itemize}
\item Fair loss link
\item Stubborn link
\item Perfect link
\item Logged perfect link
\end{itemize}

Each of these links has specific and unique characteristics.
For more information on this subject, please consult the following resources:

\begin{itemize}
\item UNIX Network Programming Volume 1, Third Edition: The Sockets Networking API By W. Richard Stevens, Bill Fenner, Andrew M. Rudoff
\item Beej's Guide to Network Programming Using Internet Sockets by Brian Hall
\end{itemize}

\section{Parallel Programming}
\label{Parallel-Programming}
In Unix, `fork()` creates a new process, where a parent process creates a new child process. The parent process must wait for the child process to exit, allowing the child process to terminate properly.
This behaviour can become problematic if the child process becomes defunct (a zombie) and the parent process ignores waiting. In some systems, the `init` process reaps (destroys) defunct processes. A child process becomes a zombie until the parent process waits or the parent ignores the `SIGCHLD` signal~\cite{brian2020, brian2015}.
Ignoring the waiting for the child process to exit in the parent process:

\begin{verbatim}
int main()
{
    signal(SIGCHLD, SIG_IGN); //don't wait
    fork();
}
\end{verbatim}

Processes and threads are core foundations for building Distributed computing systems~\cite{pcdp2020, alinush2015}.
Processes are units of work distribution.
Threads are units of concurrency.
Parallel programming presents significant challenges, including data sharing, coordination, deadlock, lock granularity, and others~\cite{pcdp2020, alinush2015}.
It is possible to have multiple threads within a single process. Some benefits of this include~\cite{pcdp2020, alinush2015}:

\begin{itemize}
\item Memory and resource efficiency due to sharing.
\item Responsiveness (no network delays within the process).
\item Performance (increased throughput); note that if the process blocks, every internal thread blocks as well.
\end{itemize}

Another combination is to have multiple processes within a node. Some benefits of this include~\cite{pcdp2020, alinush2015}:

\begin{itemize}
\item Responsiveness (mitigating JVM delays, if applicable).
\item Scalability.
\item Availability and fault tolerance.
\end{itemize}

There are different forms of parallelism (task, functional, loop, data-flow)~\cite{pcdp2020}. Java's popular fork-join framework is based on the divide and conquer paradigm, which is useful, where the problem can be decomposed in this sub-problem structure~\cite{pcdp2020}.

When using threads or processes, it is ideal to utilize the optimal number to prevent reduced performance due to degradation from increased load (due to process isolation) and communication costs for sharing information between processes.

Promises and futures are a popular form of parallelism where a deferred call is made while simultaneously performing another disjoint task in simultaneously parallel, and then joining on the result from the main thread.

Guaranteeing reproducibility can be a desirable goal. Quasi-randomness may be acceptable in some contexts. A parallel program can exhibit the following characteristics to address this behavior~\cite{pcdp2020}:

\begin{itemize}
\item Functional determinism: the same input to the same function always yields the same output.
\item Structural determinism: any repetition of the parallel program yields the same computational graph.
\end{itemize}

Let's say that your process is processing permutation-invariant data (e.g., unsorted data); in this case, only functional dependency is needed, not structural determinism. This characteristics can provide clever shortcuts to optimize your program.

\begin{figure}[H]
\centering
\includegraphics[scale=0.4]{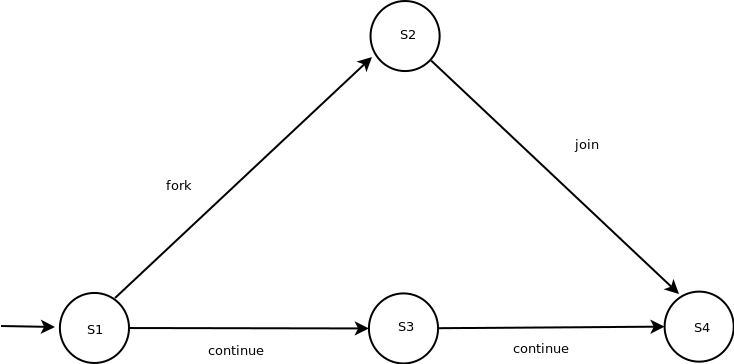}
\caption{ An example of a computational graph.}
\label{computation-graph-diag}
\end{figure}

\begin{itemize}
\item Each node represents a sequential or sub-computation step.
\item Each edge represents an ordering constraint.
\end{itemize}

Without a fork or join operation, the computational graph shows a straight line from the start to the finish nodes. In our example, S2 and S3 runs in parallel, as shown in Figure~\ref{computation-graph-diag}.

Metric performance:

\begin{itemize}
\item Work: the sum of execution times across every node.
\item Span: the length of the longest path (critical path length).
\end{itemize}

For fork-join programs, the edges can be:

\begin{itemize}
\item Continue edge: captures the sequencing of steps within a task.
\item Fork edge: connects a fork operation to the first step of a child task.
\end{itemize}

Ideal parallelism ($ip$) is given by:
$ip = work / span$
For a sequential program, the ideal parallelism is 1. This requires having a computation graph~\cite{pcdp2020}.
Unfortunately, we don't always have a computation graph. In such cases, we can use Amdahl's law, which states that the speedup of a threaded program is limited by the sequential portion of the computation across all processes~\cite{pcdp2020, alinush2015}.
There are also alternatives like share-nothing architecture (using the message-passing paradigm), e.g., actor model, and distributed actors, which make use of message passing. Non-blocking approaches using `getAndAdd()` and `compareAndSet()` are also available~\cite{pcdp2020}.
\subsubsection{Properties of a Parallel Program~\cite{pcdp2020}}

\begin{itemize}
\item Safety: bad things never happen.
\item Liveness: good things eventually happen.
\end{itemize}

Let us use the analogy of a traffic stop:

\begin{itemize}
\item Safety: vehicles can only move in one direction at a time.
\item Liveness: even with traffic, every vehicle is guaranteed to eventually leave the junction.
\end{itemize}

\subsection{Process Communication}
Fine-tuning control of computer programs involves mutual exclusion and synchronization. Let us define these two concepts:

Mutual exclusion: two or more events do not happen at the same time, e.g., event A precedes event B.

Synchronization guarantees the order of access of multiple threads to a shared resource. Synchronization outside a computer can be verified in real life by a clock. However, achieving computer synchronization without a clock is challenging, as there are also issues with the accuracy of clocks due to drifts. It is possible to do so using computational graphs, especially in a parallel program. While it is easy to know the order of execution within a process, it is challenging to ascertain the order of execution in a threaded program. Synchronization is achievable by signaling or instructing the other party to wait until a condition gets met. There are traditional synchronization problems posed in the literature with wide-applicability. These include the producer-consumer problem (with multiple variants), the reader-writer problem, the starvation problem (bounded wait on a semaphore), and the dining philosophers problem~\cite{pcdp2020, downey2016, brian2015}.

A barrier is a construct that blocks multiple threads and releases every blocked thread upon the arrival of the last one~\cite{downey2016}. This paradigm is commonly known as a phaser in Java and is widely employed in MPI for creating collective operations. This construct can also be used for pipelining~\cite{pcdp2020, downey2016, brian2015}.
A good book on interprocess communication is Beej's Guide to Unix IPC by Brian Hall. There are multiple venues for sharing information between processes in Unix.

\begin{itemize}
\item Signal: A signal is a way of performing process communication. One process raises a signal, and another process delivers it to the destination handler to effect a custom callback. Its thread-safety and interrupt-safety are uncertain.

\item Pipe: This is the simplest form of process communication, where a process writes to one end of a pipe and reads from the other end. There are many variations, such as FIFO (named pipe).

\item Message queue (`msgsnd()`, `msgrcv()`): see usage details.

\item Semaphore: At initialization, a semaphore is set to a user-defined value, representing the number of threads that can pass through the critical section before blocking. Threads can increment or decrement the semaphore's value, which cannot be read outside the semaphore construct. If the semaphore's value becomes negative, every thread will block unless it increments the value to be greater than 0. Most semaphore implementations are not atomic, and race conditions may occur if multiple threads access the resource simultaneously. One way to mitigate this issue is to have a single initialization process create the semaphore before the main processes begin to run. The main processes can then access the semaphore but cannot create or destroy it. There are similarities to restricting access using locks, permissions, etc.

\item Shared memory segments: A process writes to a memory segment, and another process reads from the same segment. Coordinated access to the segment still requires using locks or semaphores (e.g., `shmget()`, `shmat()`).
\end{itemize}

\subsection{OpenMPI}
OpenMPI~\cite{cs598-s152022, llnlgov2022, mpitutorial2022, eijkhout2022} is an implementation of the message-passing paradigm for the exchange of information between processes, widely used in rendering farms, high-performance computing, and Scientific computing. For example, Curie, a French supercomputer, makes extensive use of OpenMPI for processing workflow. One of the best resources for learning MPI on the Internet can be found on \href{https://www.open-mpi.org/}{OpenMPI website}.

The number of processes, $size$, is determined during the setup of the OpenMPI communicator. However, adjustments can be made using sophisticated process group management. Every process is assigned a rank, ranging from 0 to $size$-1.

\begin{itemize}
\item mpi\_bcast: all processes must wait until all processes have reached the same collective.
\item mpi has optimized the implementation of distributed operations.
\end{itemize}

~\textbf{Two types of communication are in use}:

\begin{itemize}
\item Point-to-Point: two processes in communication.
\item Collective: every process is communicating together
\end{itemize}

\subsubsection{Point-to-Point} 

\begin{itemize}
\item send: move data from one process to another process.
\item recv: accept data sent to the process by other processes.
\end{itemize}

\subsubsection{Collective Operation}

\begin{itemize}
\item broadcast: one process sends a message to a group of processes.
\item reduction: one process gets data from every other process and applies transformation (sum, minimum, maximum).
\item scatter: a single process partitions the data and sends each chunk to every other process e.g. MPI\_Scatter.
\item gather a single process assembly data from different processes in a buffer e.g MPI\_Gather. MPI\_AllGather is to gather and scatter the results from every process. There are synchronization primitives like locks, barrier
\end{itemize}

\subsubsection{One-Sided Communication}
MPI allows for remote memory access (RMA). Here are some commands which include:

\begin{itemize}
\item MPI\_WIN\_create()
\item MPI\_WIN\_allocate()
\item MPI\_GET()
\item MPI\_PUT()
\item MPI\_Accumulate()
\item MPI\_Win\_free()
\end{itemize}

Unlike the two-sided communication model, which requires a sender and a receiver for data transfer, the one-sided communication model allows a process to directly access another process's memory space. Consequently, only one process initiates communication directly, minimizing data transfer and CPU intervention. However, there is a caveat that the ordering of RMA operations cannot be safely guaranteed~\cite{cs598-s152022, eijkhout2022}.
In our implementation of Distributed shared primitives, we made extensive use of one-sided communication. While RMA is inherently non-blocking, we are currently employing it in a blocking mode using locks. Implementing an epoch with MPI\_Win\_fence might be a more suitable approach in our logic to fully leverage the non-blocking features. MPI\_Compare\_and\_swap could also present a better alternative~\cite{cs598-s152022, eijkhout2022}.
If one-sided communication is used in place of the default two-sided communication, the concept of explicit acknowledgments (ACKS or receipt confirmations) as in two-sided communication doesn't directly apply. The communication is more akin to direct memory manipulation.
Here is an example of an epoch:

\begin{verbatim}
MPI_Win_fence    // start of epoch
.
.
.
MPI_Win_fence    // end of epoch
\end{verbatim}

Is using a fence, better than using locks on the level of concurrency granularity? Hints, MPI\_Win\_fence is a collective using ideas from barriers.

\subsubsection{Best Practices}
\begin{itemize}
\item No user-defined operation in MPI\_Accumulate.
\item Ensure local completion before accessing the buffer in an epoch.
\item It is impossible to mix MPI\_GET, MPI\_PUT, MPI\_Accumulate in a single epoch
\end{itemize}

\subsubsection{Benefits}
\begin{itemize}
\item It can help to reduce synchronization.
\item It minimizes data movement (excluding buffering).
\end{itemize}

\chapter{Basics of Distributed Computing }
\label{Basic-Dist-Comp}
\subsubsection{ \small "Without education, we are in a horrible and deadly danger of taking educated people seriously." -- G.K. Chesterton }
\subsubsection{ \small "A distributed system is one in which the failure of a computer you didn't even know existed can render your own computer unusable." -- Leslie Lamport }

\begin{figure}[H]
\centering
\includegraphics[scale=0.67]{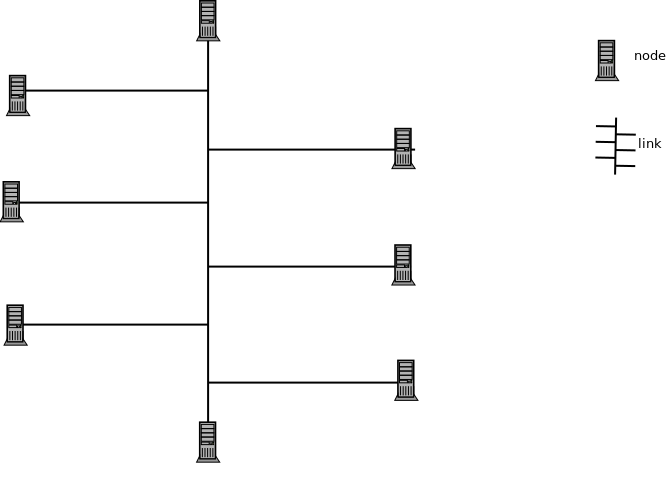}
\caption{Network of nodes in a cluster.}
\label{distributed-system-diag}
\end{figure}

Distributed Systems~\cite{pcdp2020, alinush2015} are sets of nodes (devices) connected by communication links that operate as a single, cohesive system. Although each device functions locally and independently, the overall system appears as a unified global entity, providing the "single view illusion." Examples include the Internet, edge computing environments, mobile networks, and sensor networks, as illustrated in Figure~\ref{distributed-system-diag}.

The benefits of employing a Distributed system can include:

\begin{itemize}
\item Fault tolerance (reliability)
\item Scalability
\item Load distribution
\end{itemize}

A common misconception is that a Distributed system inherently possesses scalability. However, this is not always the case. Horizontal scaling, which involves adding more nodes, yields performance gains primarily when the communication cost between nodes is low and the task at hand is parallelizable and can be effectively distributed. The challenges inherent in distributed systems often involve partial failures (such as network or node failures) and the complexities of managing concurrency.
The goal of an ideal distributed system is to provide maximum throughput with acceptable latency.
The CAP theorem (stating that only two out of three properties can be simultaneously satisfied) is an acronym used for reasoning about the fundamental trade-offs in the design of Distributed systems.

\begin{itemize}
\item C: Consistency (every read receives the most recent write or an error)
\item A: Availability (every request receives a (non-error) response, without guarantee that it contains the most recent write)
\item P: Partition tolerance (the system continues to operate despite arbitrary message loss (partitions))
\end{itemize}

Distributed Systems typically aim to be either AP (prioritizing Availability and Partition tolerance) or CP (prioritizing Consistency and Partition tolerance). Centralized systems are inherently AC (Availability and Consistency) because they do not inherently tolerate network partitions.
Peter Deutsch formulated a list of fallacies of Distributed computing, highlighting common incorrect assumptions made by developers:

\begin{itemize}
\item The network is reliable (networks experience failures; always plan for retries and acknowledgments).
\item There is zero latency (account for the non-negligible transmission time for packets across the network).
\item Bandwidth is infinite (network bandwidth is finite; therefore, segment packet sizes appropriately).
\item The network is secure (consider the security of your network, whether it is a virtual private network or a private network).
\item Network topology is fixed.
\item There is one administrator (distributed systems often require self-healing mechanisms as a single administrator might not be feasible).
\item Transport cost is zero (similar to "no latency," always anticipate delays due to network transport).
\item The network is homogeneous (expect that network partitions may occur).
\end{itemize}

When developing Distributed Software, it is crucial to avoid these oversimplified assumptions.

Consensus is achieved when every node agrees on the same value. Furthermore, the decided value must be one of the values that was initially proposed. Consensus is equivalent to atomic commits (atomic broadcast)~\cite{kthcourseds}.

Reasoning about Distributed computation can be inherently complex, necessitating the use of simplifying models. One such model involves specifying the failure assumptions, which can encompass various modes of process failure~\cite{kthcourseds}:

\begin{itemize}
\item Crash-stop: A process halts and stops sending or receiving messages.
\item Omission failures: A process fails to send or receive messages.
\item Crash-recovery: A process fails but subsequently restarts, potentially recovering its state from a backup.
\item Byzantine or arbitrary failures: A process exhibits unspecified or even malicious behavior, sending arbitrary messages or entering arbitrary states.
\item Correct processes: Processes function as intended, reliably sending and receiving messages.
\end{itemize}

In our work, we have adopted the crash-stop model due to its relative simplicity.
To achieve fault tolerance, Distributed systems must be designed to cope with failures. This is typically accomplished through a combination of techniques, including retrying lost packets, replicating data across nodes for enhanced availability, and employing mechanisms for replacing failed components (e.g., electing a new leader if the current one fails). Fault tolerance is the system's ability to maintain operation in the presence of faults.

\begin{figure}[H]
\centering
\includegraphics[scale=0.6]{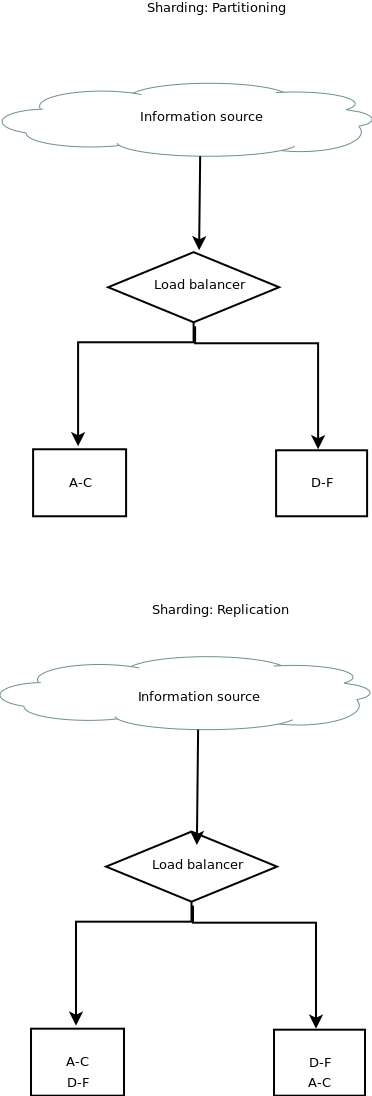}
\caption{Depiction of Sharding and Replication.}
\label{sharding-diag}
\end{figure}

One common use case for Distributed systems is Sharding, as depicted in Figure~\ref{sharding-diag}. In the context of databases, sharding is the process of horizontally partitioning a database table into multiple independent subsets. Sharding (also known as partitioning) is often combined with replication (duplicating data across multiple nodes) to enhance data availability. These techniques involve organizing data into chunks that may or may not overlap.

NoSQL database systems tend to be more readily adaptable to distributed concepts, such as key-value stores and document stores~\cite{kthcourseds, alinush2015}. Several database architectures can ensure data availability across nodes through replication strategies (e.g., master-slave, master-master, buddy replication).

A replicated state machine is a technique used to guarantee availability in the face of failures by mirroring data and state across multiple nodes. It adheres to the following fundamental properties~\cite{kthcourseds}:

\begin{itemize}
\item Every replica executes every operation.
\item The system starts in the same state across all replicas, and if all operations are deterministic, it ends in the same final state across all correct replicas.
\end{itemize}

Nodes within a Distributed system can assume various roles~\cite{petrov2019}:

\begin{itemize}
\item Coordination: Managing other nodes, such as the leader in the Raft consensus algorithm.
\item Cooperation: Participating in the execution of tasks to achieve a common goal.
\item Dissemination: Replicating data to other nodes in the system.
\item Consensus: Verifying quorums and deciding on proposed values.
\end{itemize}

A thorough understanding of network and parallel programming is a prerequisite for comprehending the intricacies of Distributed systems. We will provide a foundational overview of both these topics in Chapter~\ref{Preq-Dist-Comp}.

\section{Theoretical Foundations}

We will discuss some fundamental theoretical concepts underpinning Distributed systems, including:

\begin{itemize}
\item FLP Impossibility of Consensus
\item Two Generals Problem
\end{itemize}

\subsection{FLP Impossibility of Consensus}
The Fischer-Lynch-Paterson (FLP) impossibility result describes the inherent limitations of achieving consensus in asynchronous distributed systems. It states that in an asynchronous system subject to even a single crash-stop failure, it is impossible to guarantee that a consensus algorithm will always satisfy the following three crucial properties~\cite{kthcourseds, alinush2015, petrov2019}:

\begin{itemize}
\item Agreement: Every correct node must eventually decide on the same value.
\item Validity: If all initially proposed values are the same, then any decided value must be that value. (A weaker form states that the decided value must be one of the initially proposed values.)
\item Termination: Every correct node must eventually reach a decision.
\end{itemize}

The FLP result highlights the fundamental challenges of achieving reliable consensus in systems with unpredictable timing, applicable to asynchronous, synchronous, and partially synchronous systems~\cite{kthcourseds, alinush2015, petrov2019}:

\begin{itemize}
\item Consensus cannot be solved deterministically in a purely asynchronous system if there is even a single crash-stop failure.
\item In a synchronous system, consensus cannot be solved if $N-1$ nodes fail (where $N$ is the total number of nodes).
\item Consensus is solvable in a synchronous system with up to $t$ crash failures, provided that $2t < N$ (i.e., a majority of nodes are correct).
\end{itemize}

\subsection{Two Generals Problem}
The Two Generals Problem, also known as the Coordinated Attack Problem, illustrates the difficulties of achieving agreement over an unreliable communication channel. Two generals must agree on a time to launch a joint attack, communicating only through messengers who might be captured or delayed~\cite{kthcourseds, alinush2015, petrov2019}.
Mapping this problem to a real-world Distributed system reveals the following analogous challenges~\cite{kthcourseds, alinush2015, petrov2019}:

\begin{itemize}
\item Two nodes need to agree on a specific value before a defined time limit.
\item Communication occurs via message passing over a potentially unreliable network channel.
\item Every message ideally requires an acknowledgment to confirm receipt.
\end{itemize}

The Two Generals Problem demonstrates that it is impossible to guarantee that two parties can reach a perfect consensus in an asynchronous system when there is even a possibility of a single faulty communication link~\cite{kthcourseds, alinush2015, petrov2019}.

\section{Logical Clocks}

Time, in its ideal form, is a monotonically increasing counter with consistent intervals between increments. The duration of this interval defines the scalar units (e.g., seconds, minutes). Time becomes truly useful when it enables the ordering of events, determining precedence or simultaneity. For this ordering to be meaningful, a common reference point for initializing the counter is necessary. If clocks across different nodes are synchronized with this reference, we can accurately determine "happens-before" or concurrent relationships between events.

However, the notion of a perfectly synchronized global clock is rarely achievable in distributed systems. Therefore, it becomes crucial to determine if an event on one node was causally triggered by a previously received event from another node. This is particularly challenging in asynchronous systems, where the concept of precise timing is blurred by variable message delays across different systems. Even if individual systems possess internal clocks, synchronizing these clocks and mitigating time drift over extended periods remains a significant hurdle.

The message-passing paradigm can maintain a sense of event order by employing logical clocks, which are based on "happens-before" relationships. Logical clocks assign monotonically increasing timestamps to events to track these causal dependencies, aiming to establish a partial or total order of events in the absence of a global physical clock.

If event 'a' happens-before event 'b' on the same process, we denote this as $a \rightarrow b$, where 'a' could be a send event and 'b' a subsequent deliver event. This "happens-before" relation is fundamental for inferring event order in the absence of a global clock~\cite{kthcourseds}. Two primary types of logical clocks are commonly used:

\begin{itemize}
\item Lamport clock
\item Vector clock
\end{itemize}

To show the differences between a Lamport clock and a vector clock. We describe a distributed system comprising three processes: P1, P2, and P3. A Vector Clock is a logical time mechanism where each process maintains a vector of timestamps. The $i$-th element of the vector in process $P_j$ represents $P_j$'s knowledge of the number of events that have occurred in process $P_i$. This allows for the determination of causal relationships between events.

\subsection{Lamport clock}

The Lamport clock utilizes the "happens-before" relation to establish a total order of events. Timestamps generated by a Lamport clock have a defined less-than relationship ($<$), allowing for the ordering of any two events. However, because the timestamp is a single integer, it does not capture information about non-causality (concurrency).

\begin{figure}[H]
\centering
\includegraphics[scale=0.5]{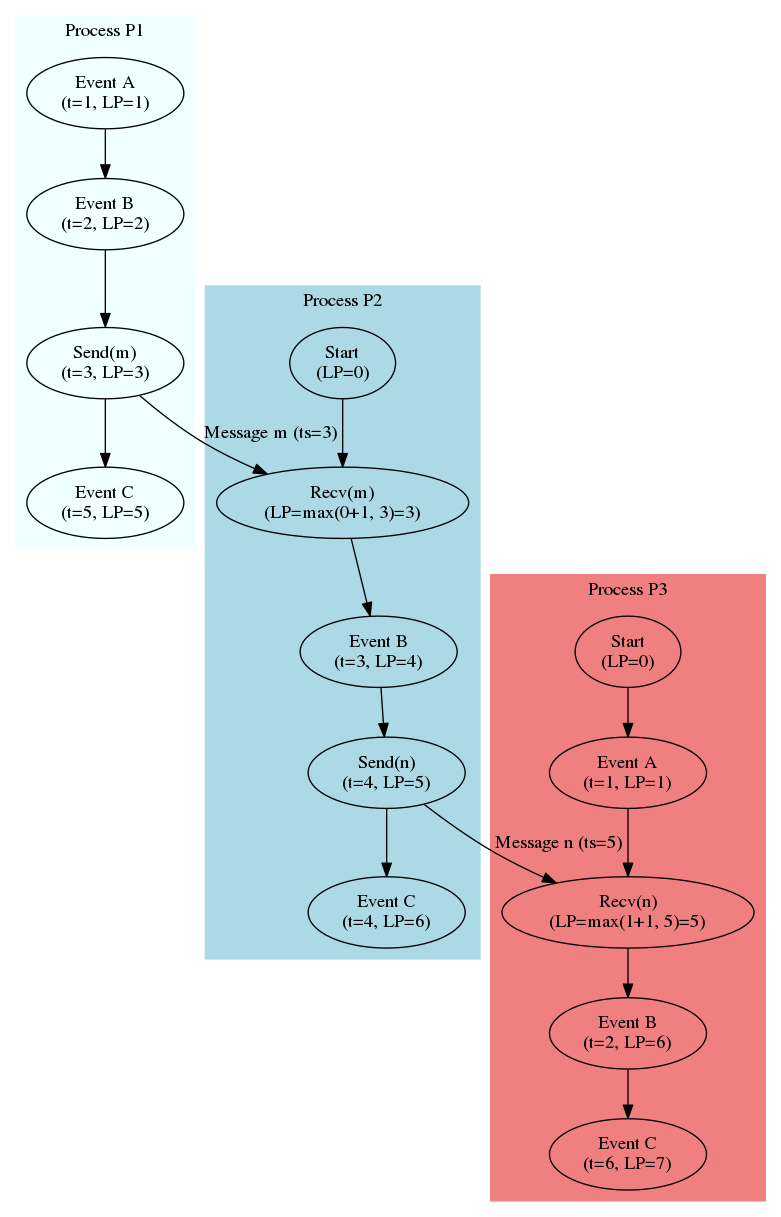}
\caption{Lamport Clock Description.}
\label{lamport_clock_diagram}
\end{figure}

The description of an example of a Lamport clock in use is shown in Figure~\ref{lamport_clock_diagram}.

\subsubsection{Process P1}
Process P1 executes a sequence of events, with its Lamport clock evolving as follows:
\begin{itemize}
    \item \textbf{Event A (LP=1)}: The initial event in P1 is assigned a Lamport timestamp of 1.
    \item \textbf{Event B (LP=2)}: The subsequent internal event increments the Lamport clock to 2.
    \item \textbf{Send(m) (LP=3)}: Before sending message $m$ to Process P2, the Lamport clock is incremented to 3. The timestamp of the send event is 3.
    \item \textbf{Event C (LP=5)}: After sending the message, an internal event C occurs, further incrementing the Lamport clock to 5.
\end{itemize}

\subsubsection{Process P2}
Process P2 starts, receives a message, executes internal events, and sends a message:
\begin{itemize}
    \item \textbf{Start (LP=0)}: Process P2 begins with a Lamport clock value of 0.
    \item \textbf{Recv(m) (LP=$\max(0+1, 3)=3$)}: Upon receiving message $m$ with a timestamp of 3 from P1, Process P2 updates its Lamport clock to the maximum of its local clock incremented by one (1) and the received timestamp (3), resulting in a timestamp of 3 for the received event.
    \item \textbf{Event B (LP=4)}: An internal event B occurs, incrementing the Lamport clock to 4.
    \item \textbf{Send(n) (LP=5)}: Before sending message $n$ to Process P3, the Lamport clock is incremented to 5. The timestamp of the send event is 5.
    \item \textbf{Event C (LP=6)}: An internal event C occurs, incrementing the Lamport clock to 6.
\end{itemize}

\subsubsection{Process P3}
Process P3 starts, receives a message, and executes internal events:
\begin{itemize}
    \item \textbf{Start (LP=0)}: Process P3 begins with a Lamport clock value of 0.
    \item \textbf{Event A (LP=1)}: An initial internal event A occurs, setting the Lamport clock to 1.
    \item \textbf{Recv(n) (LP=$\max(1+1, 5)=5$)}: Upon receiving message $n$ with a timestamp of 5 from P2, Process P3 updates its Lamport clock to the maximum of its local clock incremented by one (2) and the received timestamp (5), resulting in a timestamp of 5 for the received event.
    \item \textbf{Event B (LP=6)}: An internal event B occurs, incrementing the Lamport clock to 6.
    \item \textbf{Event C (LP=7)}: A final internal event C occurs, incrementing the Lamport clock to 7.
\end{itemize}

\subsubsection{Communication}
The interaction between processes through message passing is crucial for the Lamport clock's mechanism:
\begin{itemize}
    \item Process P1 sends message $m$ with a timestamp of 3 to Process P2, which influences the logical time of the receive event in P2.
    \item Process P2 sends message $n$ with a timestamp of 5 to Process P3, which influences the logical time of the receive event in P3.
\end{itemize}

In summary, the Lamport clock in this code snippet assigns logical timestamps to events in each process based on local increments and the timestamps of received messages, ensuring a partial ordering of events consistent with the causal "happens-before" relationship.

\subsection{Vector clock}

The Vector clock, in contrast, is based on partial order and effectively captures both causality and non-causality~\cite{attiya2004}. The ability to detect non-causality is crucial for identifying events that occur concurrently.

\begin{figure}[H]
\centering
\includegraphics[scale=0.5]{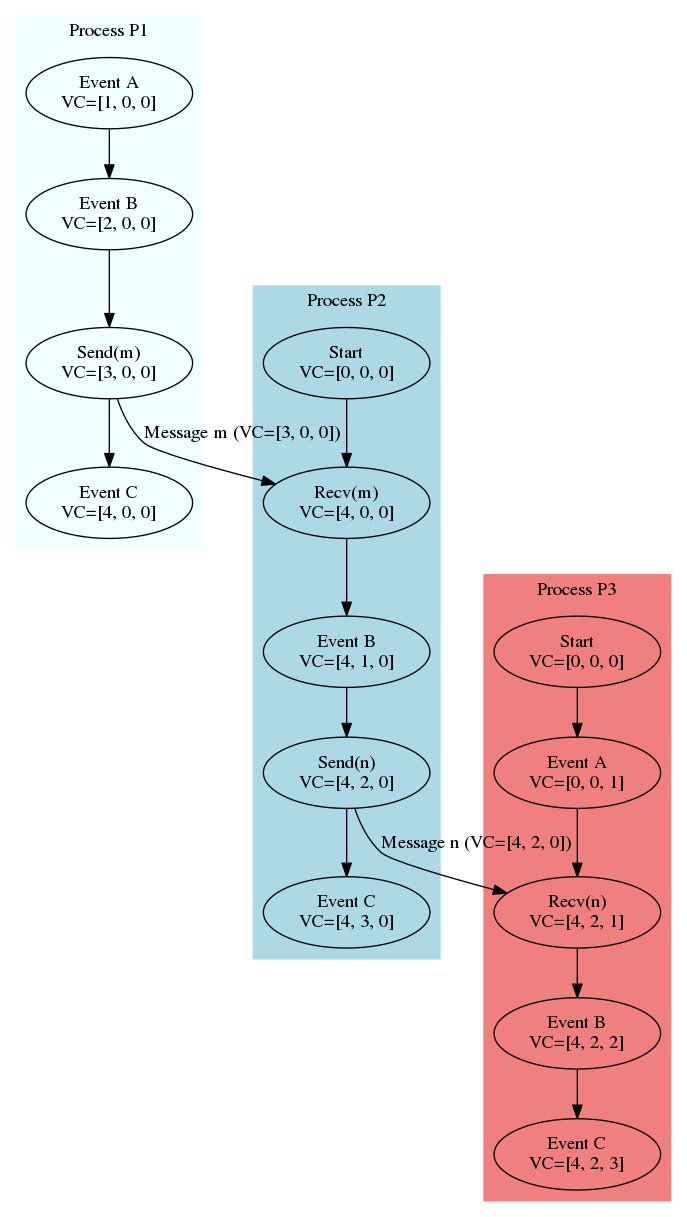}
\caption{Vector Clock Description.}
\label{vector_clock_diagram}
\end{figure}

The description of an example of a Vector clock in use is shown in Figure~\ref{vector_clock_diagram}.

\subsubsection{Process P1}
The evolution of the Vector Clock in Process P1 is as follows:
\begin{itemize}
    \item \textbf{Event A (VC=[1, 0, 0])}: Upon the occurrence of the first event in P1, the first element of its vector clock (representing its event count) is incremented to 1.
    \item \textbf{Event B (VC=[2, 0, 0])}: The second event in P1 increments its local event count in the vector clock to 2.
    \item \textbf{Send(m) (VC=[3, 0, 0])}: Before sending message $m$ to Process P2, P1's local event count is incremented to 3. The message $m$ is sent carrying this vector timestamp.
    \item \textbf{Event C (VC=[4, 0, 0])}: A subsequent internal event C in P1 increments its local event count to 4.
\end{itemize}

\subsubsection{Process P2} 
The Vector Clock in Process P2 evolves as it receives a message and executes local events:
\begin{itemize}
    \item \textbf{Start (VC=[0, 0, 0])}: Process P2 initializes its vector clock with all components set to 0.
    \item \textbf{Recv(m) (VC=[4, 0, 0])}: When Process P2 receives message $m$ with the vector timestamp [3, 0, 0] from P1, it updates its vector clock by taking the element-wise maximum of its current vector [0, 0, 0] and the received vector [3, 0, 0], and then increments its local event count (the second element), resulting in [max(0, 3)+1, max(0, 0), max(0, 0)] = [4, 0, 0].
    \item \textbf{Event B (VC=[4, 1, 0])}: A local event B in P2 increments its local event count to 1.
    \item \textbf{Send(n) (VC=[4, 2, 0])}: Before sending message $n$ to Process P3, P2's local event count is incremented to 2. The message $n$ is sent with this vector timestamp.
    \item \textbf{Event C (VC=[4, 3, 0])}: A subsequent internal event C in P2 increments its local event count to 3.
\end{itemize}

\subsubsection{Process P3} 
The Vector Clock in Process P3 evolves upon receiving a message and executing local events:
\begin{itemize}
    \item \textbf{Start (VC=[0, 0, 0])}: Process P3 initializes its vector clock with all components set to 0.
    \item \textbf{Event A (VC=[0, 0, 1])}: A local event A in P3 increments its local event count (the third element) to 1.
    \item \textbf{Recv(n) (VC=[4, 2, 1])}: When Process P3 receives message $n$ with the vector timestamp [4, 2, 0] from P2, it updates its vector clock by taking the element-wise maximum of its current vector [0, 0, 1] and the received vector [4, 2, 0], and then increments its local event count, resulting in [max(0, 4), max(0, 2), max(1, 0)+1] = [4, 2, 1].
    \item \textbf{Event B (VC=[4, 2, 2])}: A local event B in P3 increments its local event count to 2.
    \item \textbf{Event C (VC=[4, 2, 3])}: A subsequent internal event C in P3 increments its local event count to 3.
\end{itemize}

\subsubsection{Communication}
The message exchange demonstrates the propagation of vector timestamps:
\begin{itemize}
    \item Message $m$ is sent from P1 with the vector timestamp [3, 0, 0] to P2, which influences P2's vector clock upon reception.
    \item Message $n$ is sent from P2 with the vector timestamp [4, 2, 0] to P3, which influences P3's vector clock upon reception.
\end{itemize}

In summary, the Vector Clock mechanism illustrated in the code assigns a vector of timestamps to each event, allowing for the determination of causal relationships between events across the distributed system. The vector timestamps are updated based on local event occurrences and the vector timestamps received with messages, ensuring that each process maintains a view of the progress of all other processes that is consistent with causality.

A Vector clock can accurately determine whether two operations are concurrent or causally dependent on each other, a distinction that the Lamport clock cannot reliably make.

\section{Failure Detector}
A failure detector is a mechanism designed to identify faults (specifically, process crashes) in asynchronous distributed systems. The introduction of failure detectors leads to a refined classification of asynchronous systems known as "Timed Asynchronous systems" or "Partially Synchronous systems."
These systems exhibit the following characteristics~\cite{kthcourseds}:

\begin{itemize}
\item There is no guaranteed upper bound on message delivery time.
\item There is no guaranteed upper bound on process computation time.
\item However, the drift rate of local clocks is known and bounded.
\end{itemize}

A typical failure detection algorithm operates as follows~\cite{kthcourseds, alinush2015, petrov2019}:

\begin{itemize}
\item Each node in the system employs a local failure detector.
\item Initially, the failure detector's suspicions might be incorrect, but it is designed to become eventually accurate.
\item Nodes periodically exchange heartbeat messages with all other processes they believe to be alive.
\item If a node does not receive a heartbeat response from another process within a predefined timeout period, it begins to suspect that process.
\item If a message is subsequently received from a suspected node, the suspicion is typically revised (the node is no longer suspected), and the timeout period for that node might be increased.
\item Otherwise, if no communication is received after repeated timeouts, the failure detector concludes that the process has crashed.
\end{itemize}

In asynchronous systems, consensus and atomic broadcast become solvable with the aid of a failure detector.
For a failure detector to be practically useful, it must satisfy certain requirements with varying degrees of certainty~\cite{kthcourseds, alinush2015, petrov2019}:

\begin{itemize}
\item Completeness: (How promptly are crashed nodes detected?) \\*
Every process that has crashed is eventually detected by every correct (non-crashed) process (this relates to liveness).
\item Accuracy: (How often are alive nodes mistakenly suspected?) \\*
No correct process is ever suspected by any other correct process (this relates to safety).
\end{itemize}

These requirements can be further categorized as strong or weak.
In our implementations, we prioritized completeness over strong accuracy. In our specific use case, it is acceptable for a healthy process to be temporarily suspected, especially during intermittent network issues, as its status can be updated back to "alive" upon subsequent communication. We initially assumed all processes were healthy. Depending on the specific application requirements, it might be preferable to initially assume all processes are dead and require explicit confirmation of their liveness. When considering adding a timeout outside a busy-wait loop for reads, the timeout duration needs to be estimated relative to the arrival of the first message. If the arrival of the first message is not guaranteed (e.g., due to a lack of initial quorum), then the system might not function correctly from the outset.

To design a failure detector optimized for accuracy, measures are taken to minimize the suspicion of healthy processes. A common strategy is to incrementally increase the timeout period if a node fails to respond to pings, rather than immediately suspecting it as failed. This "benefit of the doubt" approach acknowledges that network communication issues might be the cause of the lack of response, rather than a process crash. By maintaining per-node timeout values and dynamically adjusting them based on communication history, the failure detector can learn the typical behavior of the network and become more resilient to transient network faults.
Failure detectors can be further enhanced by expanding the types of failures they can identify. Our implementation targets a minimal subset of potential failures. Ideally, the design of a failure detector should leverage domain-specific knowledge about the types of failures that are most likely to occur in a given system, allowing for tailored detection mechanisms.

\section{Graceful Degradation}
\label{graceful-degradation}

Distributed systems have numerous industrial applications as the Internet becomes increasingly ubiquitous. Organizations are building these large-scale services with stringent requirements for minimal downtime. Given the complexities of these interacting nodes, a myriad of issues can arise, including node failures and intermittent network calls, among others.

Graceful degradation (e.g., using a circuit breaker pattern) is a situation where, as more nodes fail, the system's functionality is reduced in a controlled manner rather than experiencing a catastrophic breakdown. This can involve reducing the components in the user interface, changing functionality as network bandwidth is throttled, or ensuring that only the most critical services remain running as the number of nodes in the cluster decreases to a bare minimum. In contrast, a quorum-based system (e.g., using Paxos or Raft) represents a form of achieving bounded graceful degradation as a fault-tolerance mechanism. These systems stop working when the number of operational nodes falls below the quorum limit.

Failure tolerance is at the core of building fault-handling mechanisms using resiliency patterns (e.g., retry, circuit breaker, fail-fast, bulkhead)~\cite{microazure}. The overarching aim of these patterns is to prevent requests from being directed to faulty resources, thereby enhancing the resilience of the distributed systems.

\begin{itemize}
\item A circuit breaker makes use of a failure detector that scans for breakdowns in your cluster. A failure is determined based on a set threshold (consider using the accuracy and completeness properties). Once a failure rate threshold is exceeded, the resource becomes unavailable and hence cannot accept new requests until a set time has elapsed, at which point it is assumed that the unavailable resource has recovered and can begin to process requests again. The recovery process is then re-triggered by a secondary resource, e.g., placing the circuit breaker behind a load balancer. When you visit a website and see a banner indicating that the site is under maintenance, you have probably been redirected by a circuit breaker in action.

\item The retry pattern would suffice for transient failures, which are short-lived errors. However, for prolonged failures, the circuit breaker is ideal. Care must be taken when using the circuit breaker to handle cascading failures, which occur when downstream systems become non-functional due to dependency on an unavailable parent system. Correlation vectors can help with distributed debugging of such failures.
\end{itemize}

It is important to consider the following recommendations when using resiliency patterns that utilize timeouts. Timeouts can hold up resources when limits are reached. Unfortunately, the risk of cascading failures on requests that are dependent on other requests can lead to a catastrophic breakdown of the system. Setting appropriate time limits for timeouts must be done in a way that delayed processes are not timed out prematurely.

\chapter{Distributed Consensus Algorithms}
\label{Dist-Con-Algo}
\subsubsection{ \small "Beware the man of a single book." -- Thomas Aquinas }

Achieving agreement among nodes is a fundamental objective that Distributed systems strive for as a core component of performing real-world tasks. Abortable consensus offers a mechanism for reaching consensus through repeated attempts across multiple rounds. The overall consensus process concludes when convergence is achieved, which occurs when a majority or all of the nodes have decided on the same value. We will discuss several well-known distributed consensus algorithms, as detailed in Sections \ref{paxos}, \ref{election}, and \ref{raft}, respectively.

\section{Paxos Algorithm}
\label{paxos}

Paxos is a fundamental algorithm for achieving consensus among multiple nodes across several rounds of communication. It provides a method for achieving abortable consensus in a distributed system with the following key properties~\cite{alinush2015}:

\begin{itemize}
\item Single point of failure.
\item Accommodation of network partitions.
\end{itemize}

The algorithm generally follows these steps~\cite{kthcourseds, petrov2019}:

\begin{itemize}
\item Each node in the system is assumed to have a network channel enabling communication with every other node.
\item Only designated proposers can suggest values to be decided by the consensus algorithm.
\item Upon reaching consensus, every node in the system will have decided on the same value.
\end{itemize}

The following concurrency-related properties~\cite{kthcourseds, alinush2015, petrov2019} are generally applicable in parallel and concurrent programming paradigms:

\begin{itemize}
\item Liveness: A majority of the nodes must be able to reliably communicate to exchange values and make progress.
\item Safety: Only values proposed by proposers can be chosen as the consensus value.
\end{itemize}

The main characteristics of the Paxos algorithm are~\cite{kthcourseds, petrov2019, alinush2015}:

\begin{itemize}
\item Abortable consensus: Multiple rounds of communication might be necessary before a value is agreed upon.
\item Agreement based on a quorum: A quorum (typically a majority) of nodes must decide on a value for it to be considered the consensus value.
\end{itemize}

The Paxos algorithm defines several roles that nodes in the system can assume:

\begin{itemize}
\item Client: The entity that initiates the consensus process by proposing a value.
\item Proposer: A node that receives a proposal from a client and attempts to get it accepted by the acceptors.
\item Acceptor: A node that receives proposals and votes from proposers, deciding whether to accept a proposed value.
\item Learner: A node that learns the decided value once a quorum of acceptors has accepted it.
\end{itemize}

We will now discuss Single Value Paxos and Sequence Paxos in the subsequent subsections.

\subsubsection{Single Valued Paxos}

The Single Valued Paxos algorithm proceeds as follows~\cite{kthcourseds, alinush2015}:

\begin{itemize}
\item A client sends a proposed value to one or more proposers.
\item Proposers send prepare requests (containing a proposal number) to a quorum of acceptors.
\item Proposers receive responses from the acceptors. If a quorum of acceptors responds, the proposer proceeds to the next phase.
\item Proposers send accept requests (containing the proposal number and a proposed value) to the same quorum of acceptors. The value might be the original proposed value or the highest-numbered value already accepted by any of the acceptors in their responses to the prepare request.
\item Once a quorum of acceptors has accepted an accept request for a particular value, that value is considered chosen.
\item The chosen value is then communicated to the set of learners.
\end{itemize}

Properties of the Paxos algorithm~\cite{kthcourseds, alinush2015}:

\begin{itemize}
\item Validity: If a value is decided, it must have been proposed by some client.
\item Uniform agreement: No two processes decide on different values.
\item Integrity: Each process can decide on at most one value.
\item Termination: Every correct process eventually decides on a value, assuming a majority of processes are correct and reliable communication exists.
\end{itemize}

\begin{figure}[H]
\centering
\includegraphics[scale=0.52]{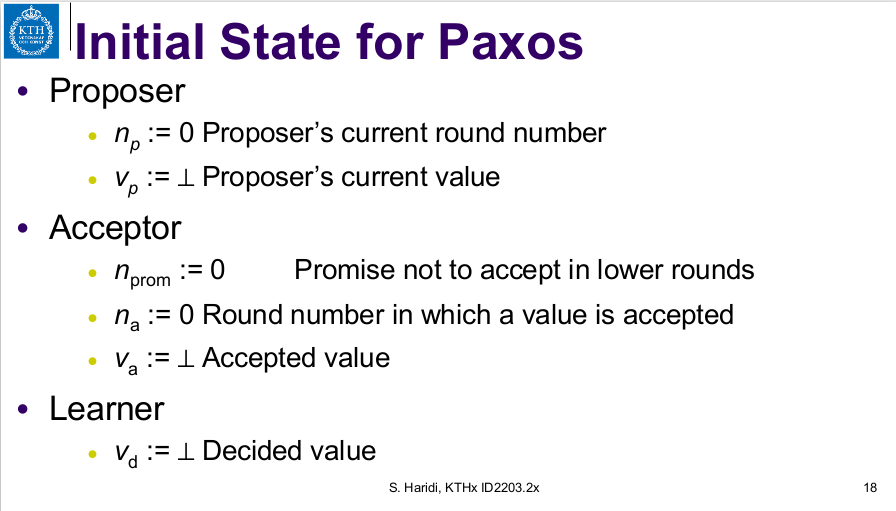}
\caption{Initialization of states at the internal start of Paxos Algorithm~\cite{kthcourseds}.}
\label{paxos1-diag}
\end{figure}

\begin{figure}[H]
\centering
\includegraphics[scale=0.52]{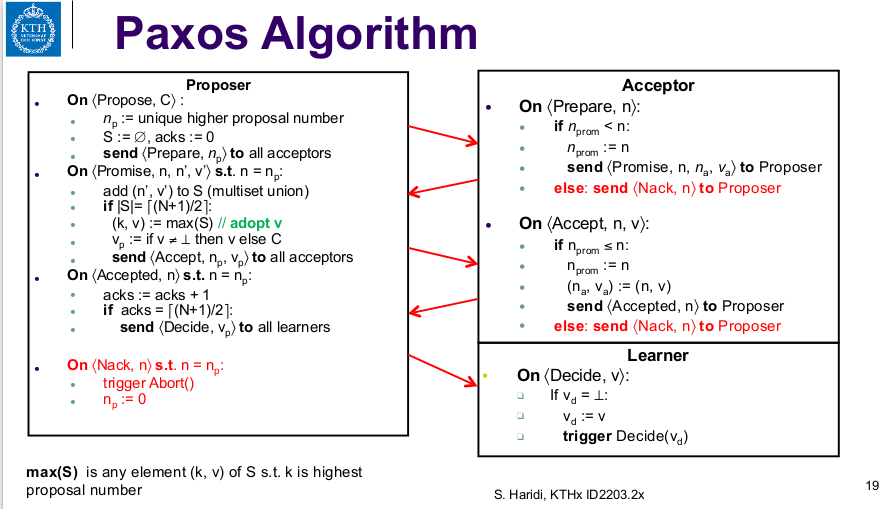}
\caption{Sequence of message flow in a Paxos Algorithm~\cite{kthcourseds}.}
\label{paxos2-diag}
\end{figure}

The implementation of the Single Value Paxos algorithm discussed in this book is visually represented in Figure~\ref{paxos1-diag} and Figure~\ref{paxos2-diag}. Figure~\ref{paxos1-diag} illustrates the initial states of each node at the internal start of the Single Value Paxos process.

Figure~\ref{paxos2-diag} presents a sequence (interaction) diagram depicting the messages exchanged between the different roles (Client, Proposer, Acceptor, Learner) in the network during the Single Value Paxos algorithm.

It's worth noting that there are several variations of Single Value Paxos, including Fast Paxos, Egalitarian Paxos, and Flexible Paxos~\cite{petrov2019}, each offering different performance characteristics or fault-tolerance trade-offs.

\subsubsection{Sequence Paxos}

The Sequence Paxos algorithm extends the Single Value Paxos to achieve agreement on a sequence of values (often representing a log of events) in a consistent order across all nodes~\cite{kthcourseds, petrov2019, alinush2015}. The general approach involves:

\begin{itemize}
\item Repetitively performing the Single Value Paxos algorithm for each position (slot) in the sequence, ensuring that the decisions for each slot are made in order.
\item Maintaining a strict prefix invariant: If all nodes have decided on the values for the first $i$ slots, then these decided prefixes must be identical across all nodes.
\item Effectively implementing an ordered atomic broadcast mechanism, where all messages are delivered to all correct processes in the same order.
\end{itemize}

In more detail, the fundamental principles of the Sequence Paxos algorithm include:

\begin{itemize}
\item Achieving agreement on a common ordering of events or operations across all participating processes.
\item Constructing a consistent log of decided values by running multiple instances of the Single Value Paxos algorithm, one for each position in the log.
\end{itemize}

\begin{figure}[H]
\centering
\includegraphics[scale=0.52]{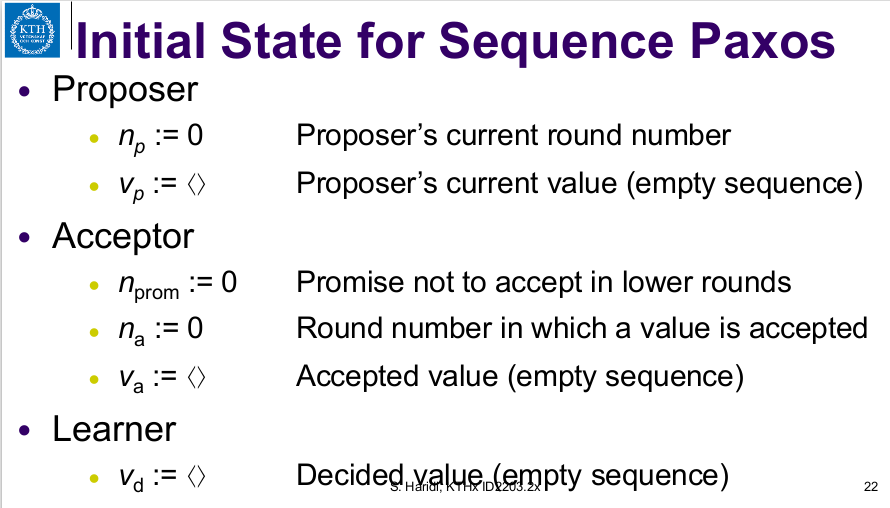}
\caption{Initialization of internal states at the start of Sequence Paxos Algorithm~\cite{kthcourseds}.}
\label{sequence-paxos1-diag}
\end{figure}

\begin{figure}[H]
\centering
\includegraphics[scale=0.52]{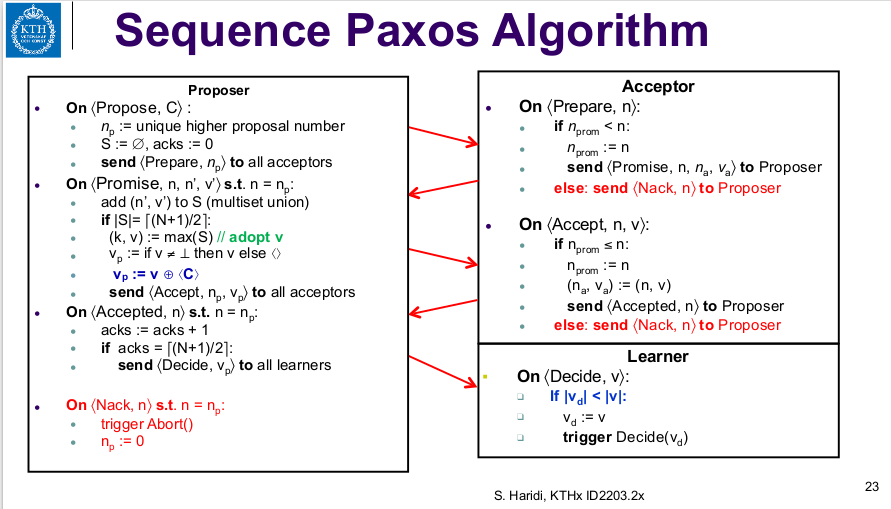}
\caption{Message flow in a Sequence Paxos Algorithm~\cite{kthcourseds}.}
\label{sequence-paxos2-diag}
\end{figure}

\begin{figure}[H]
\centering
\includegraphics[scale=0.52]{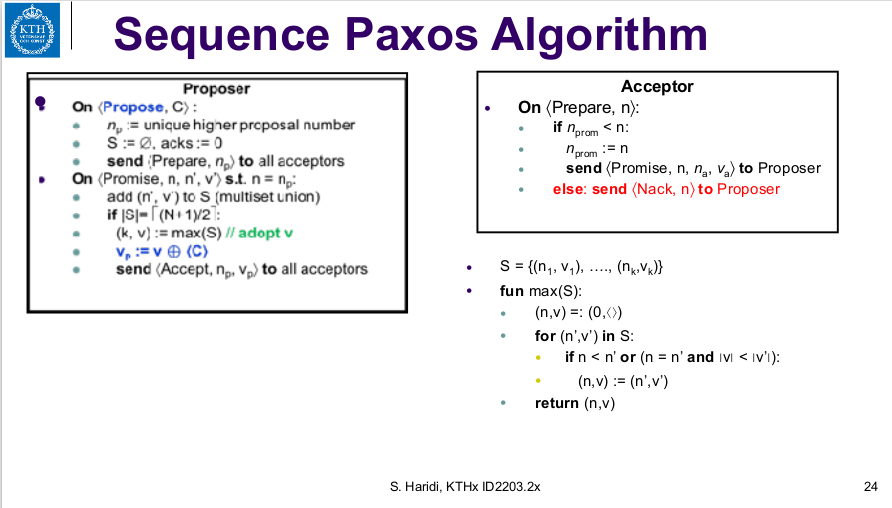}
\caption{Maintaining prefix of log invariants in a Sequence Paxos Algorithm~\cite{kthcourseds}.}
\label{sequence-paxos3-diag}
\end{figure}

The implementation of the Sequence Paxos algorithm discussed in this book is illustrated in Figure~\ref{sequence-paxos1-diag}, Figure~\ref{sequence-paxos2-diag}, and Figure~\ref{sequence-paxos3-diag}.

Figure~\ref{sequence-paxos1-diag} shows the initial internal states of each node at the beginning of the Sequence Paxos process.

Figure~\ref{sequence-paxos2-diag} presents a sequence (interaction) diagram illustrating the message flow between the different roles in the network during the Sequence Paxos algorithm.

Figure~\ref{sequence-paxos3-diag} highlights the algorithmic adjustments and mechanisms employed to maintain the crucial prefix invariants in the log as the Sequence Paxos algorithm progresses.

\section{Election Algorithm }
\label{election}

In distributed systems, allowing multiple nodes to propose values concurrently can introduce complexities in coordinating these proposals and can lead to increased contention for system resources. To streamline the consensus process and minimize such conflicts, it is often advantageous to designate a single node as the leader. This leader then assumes the role of a central coordinator, responsible for initiating and managing the agreement process. Our approach to leadership election relies on an underlying failure detector. This is a crucial component that continuously monitors the health and responsiveness of the participating nodes to identify potential failures or unresponsiveness of the currently designated leader~\cite{kthcourseds, petrov2019}. (For a more detailed explanation of failure detectors, please refer to the preceding section.)

We are using a variation of the bully algorithm which is described below~\cite{petrov2019}.
The algorithm does the following:

\begin{itemize}
\item A server can become a candidate by waiting for some set time and if no ping from others is received. It becomes a candidate.\\*
Note: timeout must be longer than the duration of the leader election. \\*
The choice of delay has a significant impact on leader election. \\*
The rule of thumb for deciding the duration
    \begin{itemize}
    \item If the value is set too low, then the second candidate begins the election before the first election triggered by the first candidate.
    \item if too high, then it will take too long for the election to start after the old leader has died. The new candidate starts an election.
    \end{itemize}
\end{itemize}

In summary, the algorithms can be described as follows: When a server suspects that the current leader has failed or become unresponsive (this suspicion typically arises from the absence of expected communication, such as periodic heartbeat messages, within a predefined timeout period, as determined by the failure detector), it initiates an election process. The goal of this election is to select a new leader from the set of currently active and available servers.

\begin{figure}[H] 
\centering
\includegraphics[scale=0.5]{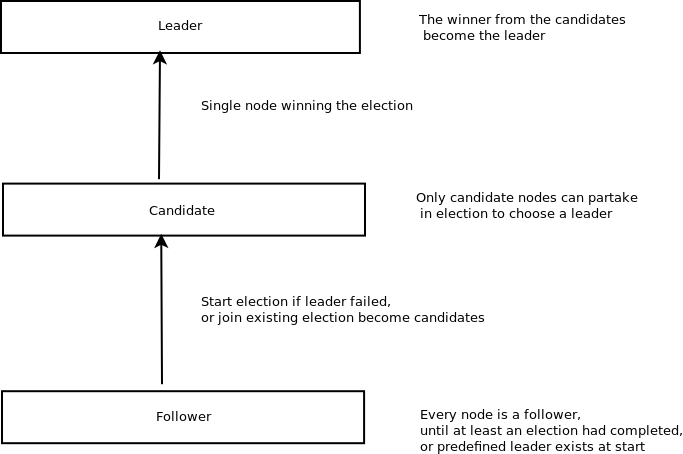}      
\caption{Depiction of Leader Election. }
\label{leader-election-diag}
\end{figure}

Here is a diagram showing how leader election works as shown in Figure~\ref{leader-election-diag}.

A variation using ballots for a leadership election to choose a leader from a set of candidates is shown below. The following code snippet presents a ballot-based approach for electing a leader from a set of candidate servers within an MPI (Message Passing Interface) environment as seen in \href{https://github.com/kenluck2001/DistributedSystemReseach/blob/master/textbook/leader-election3.c}{leader-election3.c}:

\begin{verbatim}
// Function responsible for initiating and managing the leader election process.
void leaderElection(int my_rank, int num_procs, int *leader, 
MPI_Datatype mpi_data_type) {
    process_cnts = create_shared_var(MPI_COMM_WORLD, 0);
    int cnt = 0; // Local counter for the current election attempt
    int run_loop = 1;
    // Example timeout duration for detecting leader failure
    double duration = 2.0; 
    double beg = MPI_Wtime();
    double end = beg;
    int msg_flag;
    MPI_Status status;
    int is_leader_dead = (*leader == MPI_PROC_NULL); 
    // Initially true if no leader is known
    MPI_Request requests[num_procs];

    while (run_loop) {
        is_leader_dead = isLeaderFailed((*leader));
        if (is_leader_dead) {
            beg = MPI_Wtime();
            while ((end - beg) <= duration) {
                end = MPI_Wtime();
                MPI_Iprobe(MPI_ANY_SOURCE, MPI_ANY_TAG, 
MPI_COMM_WORLD, &msg_flag, &status);
                // Break the timeout loop if any message is received
                if (msg_flag) break; 
            }
            // Check again if another process has already started an election
            // and potentially updated the leader.
            is_leader_dead = isLeaderFailed((*leader));
            if (is_leader_dead) {
                printf("Process %d: A leader process (pid=%d) 
has failed or no initial leader\n", my_rank, *leader);
                data package;
                memset(&package, 0, sizeof(package));
                // Generate a unique ballot based on time
                package.ballot = (int)time(NULL); 
                package.pid = my_rank;

                printf("Process %d: An election has been triggered 
with ballot %d\n", my_rank, package.ballot);
                beginElection(package, num_procs, requests, mpi_data_type);
            }
        }
        // Increment the shared counter to track the number of processes
        // that have reached this point in the election cycle.
        cnt = increment_counter(process_cnts, 1);
        // Determine if a quorum of processes has participated in this round.
        if (cnt == (int) MAX(((num_procs + 1) / 2.0), 1)) {
            run_loop = 0; // Exit the election loop once a quorum is reached.
        }
    }
    if (process_cnts != NULL && process_cnts->win != MPI_WIN_NULL) {
        SAFE_MPI_CALL(MPI_Win_free(&process_cnts->win));
        free(process_cnts);
    }
}

// Function for a process to check incoming election messages and potentially
// update the current leader based on the received ballot.
void checkLeader(data recv, int my_rank, int num_procs, int *leader, 
MPI_Request requests [], MPI_Datatype mpi_data_type) {
    // Static local variable to track the highest ballot seen
    static int max_ballot = -10000; 
    // Static local counter for the number of promises received
    static int cnt = 0;             
    enum msgTag ctag;
    data accept_value;
    memset(&accept_value, 0, sizeof(accept_value));

    // If the received ballot is higher than the current maximum, update it.
    if (recv.ballot > max_ballot) {
        max_ballot = recv.ballot;
        // Store the data associated with the highest ballot.
        accept_value = recv; 
    }

    // Increment the shared counter for the number 
    // of processes that have responded.
    cnt = increment_counter(proc_cnts, 1);
    int promise_cnt = (int) MAX(((num_procs + 1) / 2.0), 1); // Quorum size.

    // If a quorum of processes has responded and the current leader is not
    // the process with the highest ballot, then broadcast the new leader.
    if (cnt == promise_cnt && (*leader != accept_value.pid)) {
        ctag = mSetLeader;
        printf("Process %d: Sending broadcast to set leader to %d 
with ballot %d\n", my_rank, accept_value.pid, accept_value.ballot);
        *leader = accept_value.pid; // Update the local leader.
        for (int other_rank = 0; other_rank < num_procs; other_rank++) {
            SAFE_MPI_CALL(MPI_Isend(&accept_value, 1, 
mpi_data_type, other_rank, ctag, MPI_COMM_WORLD, &requests[other_rank]));
        }
    }
}
\end{verbatim}

\subsubsection{Note}

Remember to reset both counters for both functions at the end of the leader election.

It is crucial to remember to reset the shared counters (`process\_cnts`, `proc\_cnts`, and `process\_max\_ballot`) and to release their associated MPI Window objects using `MPI\_Win\_free()` at the end of the leader election process. Failure to do so can lead to interference with subsequent MPI operations. The provided `resetElectionCounters()` function attempts to perform this cleanup. Additionally, you will need to free the MPI datatype `mpi\_data\_type` using `MPI\_Type\_free()`.

The `MPI\_Barrier(MPI\_COMM\_WORLD)` calls are essential for ensuring that all processes synchronize at specific points within the algorithm. In the `leaderElection()` function, the barrier ensures that all processes have completed their initial leader failure detection and potential election triggering before proceeding to the next iteration of the loop. In the `main()` function, barriers are used to synchronize the printing of leader information and to ensure that all processes participate in the `checkLeader()` phase after receiving potential leader announcements.

The provided code snippet includes placeholders and assumes the existence of atomic primitives (`create\_shared\_var`, `increment\_counter`, `reset\_var`) that would typically be implemented using MPI Windows for shared memory access and synchronization, as potentially demonstrated in your Paxos implementation. You will need to ensure that these primitives are correctly implemented and accessible within your election algorithm code. The `isLeaderFailed()` function is a simplified placeholder for failure detection and should be replaced with a more robust mechanism (e.g., monitoring heartbeats) in a production-ready system. The `checkLeader()` function now uses static local variables (`max\_ballot` and `cnt`) to maintain state across multiple invocations within a single election cycle. The `initializeElectionCounters()` function is added for the proper initialization of the shared counters and variables before the election begins. The `SAFE\_MPI\_CALL` macro is included as a good practice for handling potential errors in MPI function calls. The `main()` function provides a basic example of how the `leaderElection` and `checkLeader` functions might be used within an MPI program.

Add the required barriers to make the example work as expected. Look at my implementations of Paxos on how to create proper atomic primitives.

\section{Raft Algorithm}
\label{raft}

The Paxos algorithm, while foundational for achieving consensus in distributed systems, is often perceived by software engineers as complex and challenging to reason about. This complexity motivated the invention of the Raft algorithm, designed to be a more understandable alternative while providing equivalent functionality. Raft is described as a leader election-based sequence Paxos, essentially integrating the concepts of Paxos, a replicated log, and a designated leader~\cite{kthcourseds, alinush2015}. The elected leader acts as the sole proposer, centralizing coordination and minimizing potential contention. To ensure fault tolerance, Raft employs leader election, which is triggered if the current leader is detected as dead.

The Raft algorithm defines three distinct roles for the participating nodes:

\begin{itemize}
\item Candidate: A node that initiates an election with the goal of becoming the leader.
\item Leader: The candidate that successfully wins the election and assumes the responsibility of coordinating the consensus process.
\item Follower: A participant in the Raft cluster who is not currently a candidate or a leader. Followers passively listen for instructions from the leader and vote in elections.
\end{itemize}

Despite its design for simplicity, Raft can still encounter potential issues:

\begin{itemize}
\item Multiple leaders: Although Raft is designed to have a single leader, scenarios such as network partitions or subtle timing issues during leader election could theoretically lead to the temporary existence of multiple nodes believing they are the leader. This can result in conflicting proposals and compromise the correctness of the consensus algorithm.
\item No leader: Problems during the leader election process, such as unhandled ties in votes or persistent faults affecting potential candidates, can lead to a situation where no leader is successfully elected. Without a leader, the Raft cluster cannot make progress on proposing and committing new log entries.
\item Missing log entries: Unexpected errors, such as node crashes or network disruptions, could potentially cause some decided log entries to be lost or not fully replicated across the cluster.
\item Divergent logs: If the cluster experiences issues like multiple leaders (even transiently) or failures during log replication, different nodes might end up with logs containing conflicting decided values at the same index. This divergence violates the fundamental consistency property that Raft aims to maintain.
\end{itemize}

Let us consider the following questions in the context of Raft:

\begin{itemize}
\item Can an election choose multiple leaders~\cite{alinush2015}? Yes, although Raft's mechanisms are designed to strongly discourage this, transient network issues or timing-dependent failures could theoretically lead to split votes and the possibility of multiple nodes believing they are leaders for a short period.
\item Can an election fail to choose a leader~\cite{alinush2015}? Yes, scenarios such as all candidates failing or a persistent split vote across the cluster could result in no single candidate obtaining a majority, thus leading to a failed election.
\end{itemize}

In our implementation of leader election (as discussed in the previous section), we incorporated careful delay management and election logic to significantly reduce the practical likelihood of these issues occurring. However, the theoretical possibility remains, especially in challenging network environments.

As noted, if both the answers to the questions above are true (multiple leaders or no leader), then the risk of divergent logs becomes considerably higher.

Raft shares significant similarities with both leader election mechanisms and sequence Paxos. Given that we have already implemented both of these foundational components, implementing the Raft algorithm by combining their functionalities becomes a relatively straightforward task. The following pseudocode snippets illustrate how this combination could be structured, resembling actual source code organization:

\textbf{Algorithm 1:} Basic role assignment based on the elected leader.

\begin{verbatim}
if (rank == leader)
{
    // This node acts as the proposer, 
    // initiating and coordinating log replication.
}
else
{
    // Example: Assign nodes with even rank as acceptors.
    isAcceptor = rank % 2; 
    if (isAcceptor)
    {
        // This node acts as an acceptor, voting on proposed log entries.
    }
    else
    {
        // This node acts as a learner, receiving 
        // and storing committed log entries.
    }
}
\end{verbatim}

\textbf{Algorithm 2:} Demonstrating an alternative grouping strategy with a single acceptor.

\begin{verbatim}
if (rank == leader)
{
    // This node acts as the proposer.
}
else
{
    // Example: The node with rank immediately 
    // following the leader is the learner.
    isLearner = (leader + 1) % n; 
    if (isLearner)
    {
        // This node acts as the learner.
    }
    else
    {
        // This node acts as the acceptor. In this example, 
        // there is only one acceptor (excluding the leader and learner).
    }
}
\end{verbatim}

These algorithms demonstrate the flexibility in assigning roles within a distributed consensus system once a leader has been elected. The specific grouping and role assignment strategies can be tailored based on the desired fault tolerance characteristics and performance considerations of the Raft implementation. As suggested, the possibilities are limited only by our imagination in how we structure the roles of acceptors and learners once a stable leader has been established through the leader election process.


\section{Stabilization Algorithm }
\label{stabilization}

This algorithm permits the return of a distributed computation to a 'correct behavior' from a perturbed state~\cite{lynch1993}. This perturbation may be due to failures, such as those affecting networks and nodes. Self-repairing algorithms guarantee correct behavior even in the presence of failures~\cite{tel2000}. Robust algorithms, as discussed in Chapter~\ref{Dist-Con-Algo}, have a fixed bound on the number and types of failures they can tolerate while exhibiting graceful degradation. In contrast, stabilization algorithms allow for any number of failures, with an expected delay before incorrect operation may occur until the failure is repaired and the system returns to a correct state. Stabilization algorithms are ideal for transient errors, as discussed in Section~\ref{graceful-degradation}. Most robust algorithms follow a crash-stop model, whereas stabilization algorithms follow a crash-recovery model. Stabilization can be achieved through several means, including eliminating illegal states (domain restriction) and snapshotting.

Eliminating illegal states can be illustrated with the example of a mobile phone's torchlight. There is a possibility that the torchlight might be accidentally switched on, potentially draining the battery. This condition can be avoided by redesigning the interface to require the user to continuously press a button to activate the light; releasing the button would switch the light off. Thus, the illegal state of the light being indefinitely switched on is eliminated.

Snapshotting involves saving the configuration states of a distributed computation for later retrieval~\cite{tanenbaum2007, tel2000}. When an error is detected, the system can revert to a previously known good configuration saved in the snapshot (a change point). Snapshotting is desirable when manual intervention to resolve errors is costly. Retrieving the saved change point when an illegal state is reached can guarantee self-healing in the distributed computing system. For reliability, it is essential that the saved change point has not been compromised by a malicious node, ensuring the system's return to correct behavior.

We have implemented a mechanism for saving the decided value in a single-valued Paxos algorithm. The following outline is a snippet from the file named \href{https://github.com/kenluck2001/DistributedSystemReseach/blob/master/textbook/single-paxos3-snapshot.c}{single-paxos3-snapshot.c}. This implementation includes some custom details that are not part of the general Paxos specification. Specifically, I created a new role, 'telemetry,' in addition to the standard proposer, acceptor, decider, and learner roles. I also created a dedicated communicator for the telemetry role to simulate logging the distributed system's state as quickly as possible. This is a simplification, as we have omitted the data storage logic.

\begin{verbatim}
enum manageTag {mSNAPSHOT, mREVERT};

void trigger_snapshot(MPI_Comm comm, data payload_backup,
 int num_procs, MPI_Datatype mpi_data_type, MPI_Request requests[]) {
    enum manageTag ctag = mSNAPSHOT;
    for (int other_rank = 0; other_rank < num_procs; other_rank++)
    {
        MPI_Isend(&payload_backup, 1, mpi_data_type, other_rank, 
ctag, comm, &requests[other_rank]);
    }
}

void handle_snapshot_messages(MPI_Comm comm, int my_rank, 
MPI_Request requests[], MPI_Status status[], 
MPI_Datatype mpi_data_type, 
struct mpi_counter_t *telemetry_msg_cnts) {
    int flag = -1;
    int cnt =0;
    int ret;
    data recv;
    memset(&recv, 0, sizeof(recv));

    while (1)
    { 
        /* Receive message from any process */
        if(flag != 0)
        {
            ret = MPI_Irecv(&recv, 1, mpi_data_type, 
MPI_ANY_SOURCE, MPI_ANY_TAG, comm, &requests[my_rank]);

            flag = 0;
        }
        MPI_Test(&requests[my_rank], &flag, &status[my_rank]);

        if (flag != 0)
        {
            if (ret == MPI_SUCCESS )
            {
                enum manageTag tag = status[my_rank].MPI_TAG;
                int source = status[my_rank].MPI_SOURCE;

                if (tag == mSNAPSHOT)
                {
                    printf("################ SAVING SNAPSHOT ###############");
                    printf("recv.custom_round_number: %d, 
recv.round_number: %d, recv.value: %d\n", recv.custom_round_number, 
recv.round_number, recv.value);
                    printf("################# END SNAPSHOT ################");

                }
                else if (tag == mREVERT)
                {
                    printf("################ DELETE SNAPSHOT ###############");
                    printf("PERFORM CUSTOM LOGIC FOR MANAGING SNAPSHOT");
                    printf("################# END SNAPSHOT ################");
                }

                cnt = increment_counter(telemetry_msg_cnts, 1);
            }
            flag = -1;
        }

        if (cnt>0)
        {
            if (!flag)
                MPI_Cancel( &requests[my_rank] );
                break;
        }          
    }
}

// trigger a snapshot to saved data known as recv
trigger_snapshot(row_comm[TELEMETRY], recv, num_procs, 
mpi_data_type, requests);

// retrieve snapshot 
handle_snapshot_messages(row_comm[TELEMETRY], my_rank, 
requests, status, mpi_data_type, telemetry_msg_cnts);
\end{verbatim}

To run the program

\begin{verbatim}
$ mpicc single-paxos3-snapshot.c && mpiexec -n 5 ./a.out
\end{verbatim}

\section{Byzantine Protocol (BFT) }
\label{PBT}

The resiliency discussed in SubSections~\ref{paxos}, \ref{raft} would fail if there is $< \frac{1}{2}$ of the number of live processes. More robust algorithms can handle cases of failure exceeding the tolerance bound of traditional Paxos and raft algorithms. Some algorithms have mechanisms to handle cases where the failed process can take arbitrary actions such as sending bad messages and activating unnecessary state transitions, and even denial of service can affect a distributed system. It does not prevent some interruption, but it can mitigate the attack and ensure that an agreement can still be reached.

These algorithms with robust error handling included in the consensus algorithm are known as Byzantine protocols. These allow for building distributed systems that function in chaotic channels with significant disruption, as we can achieve a quorum with at least $\frac{1}{3}$ of the number of live processes. We consider scenarios with parallel leaders and also scenarios with a single leader serving as the proposer. There are several papers describing BFT~\cite{CastroL02, Stathakopoulou2022} with varying tolerance guarantees.


\section{Distributed Commit and Transaction}
\label{transactional}

Commit is the immutable operation that finalizes a transaction, making its modifications permanent and visible throughout the system. In contrast, a transaction represents a logical and atomic unit of work, potentially comprising multiple individual operations (reads, writes, updates, deletions). A transaction is the combined procedure for performing a task, consisting of a sequence of steps. Commit serves as the concluding step of a transaction, confirming its successful execution and guaranteeing the durability of all associated changes. Transactions prevent race conditions, inconsistent updates, and conflicting operations, and can provide strict guarantees on the execution plan.

Atomic commit requires that a collection of processes jointly decide whether a transaction is committed or aborted. The properties of a transaction include:
\begin{itemize}
\item Atomicity: a transaction is either fully committed or rolled back.
\item Consistency: the last write is the latest read.
\item Isolation: no interaction between processes.
\item Durability: committed transactions are saved for later access.
\end{itemize}

We discuss two categorizations of transactions, as shown below:
\begin{itemize}
\item Flat transaction is a set of sequential actions synchronized across nodes.
\item Nested transaction has a hierarchical structure where the transaction can trigger a set of sub-transactions.
\end{itemize}

\begin{figure}[H] 
\centering
\includegraphics[scale=0.4]{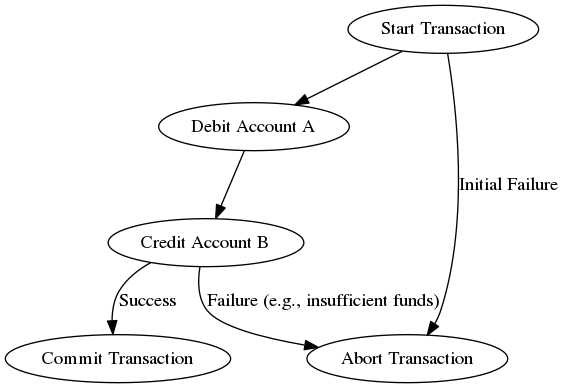}      
\caption{Transaction showing deposit and withdrawal. }
\label{flat_banking_transaction}
\end{figure}

\begin{figure}[H] 
\flushleft
\includegraphics[scale=0.42]{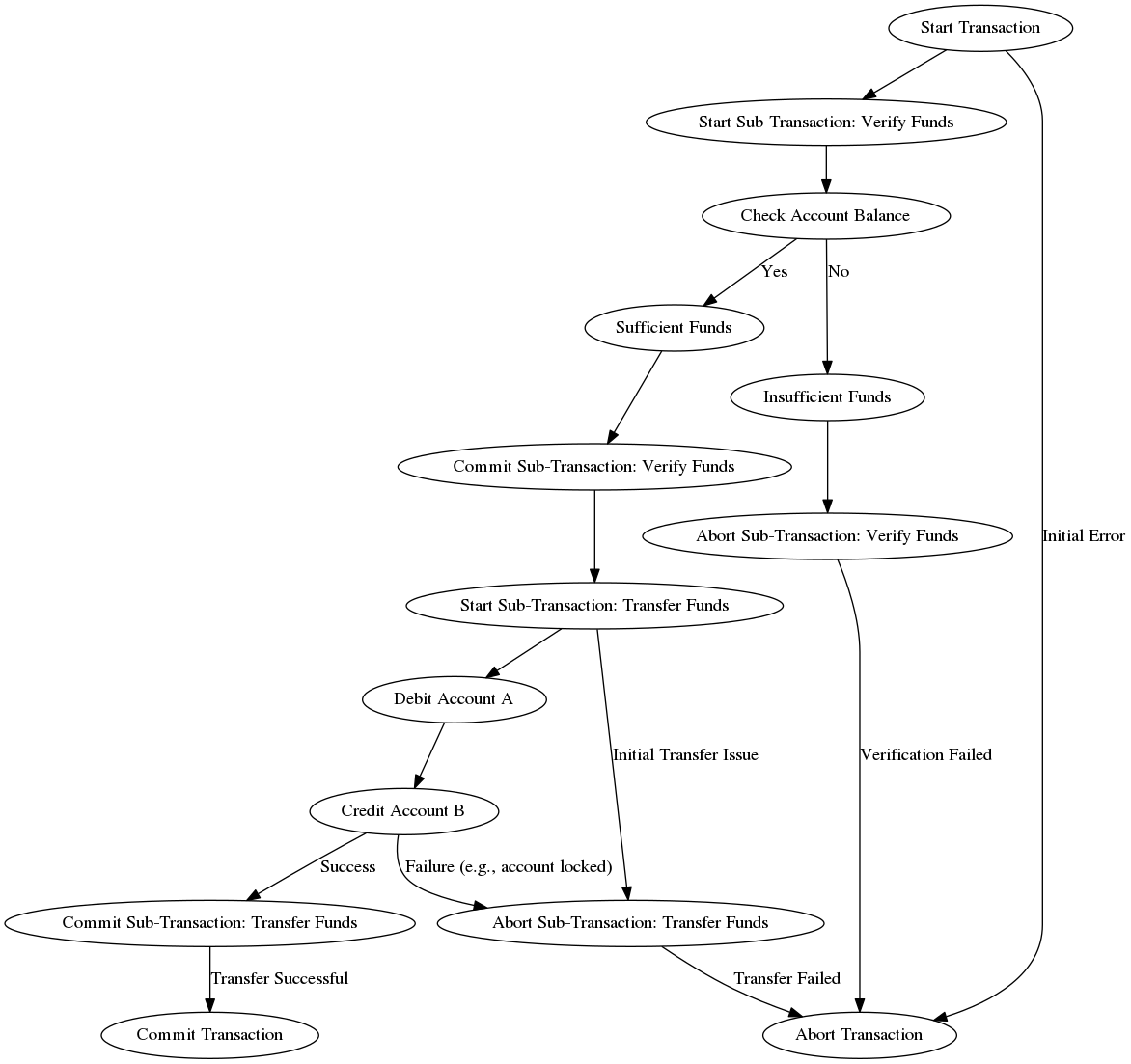}      
\caption{Nested Transaction containing sub-transactions depicting a bank transaction. }
\label{nested_banking_transaction}
\end{figure}

Let's set up a simple, single-system banking scenario with two accounts (A and B) to prepare for our discussion of distributed protocols in SubSection~\ref{Atomic-Commit-Protocols}. We will illustrate a basic transfer using a flat transaction (Figure~\ref{flat_banking_transaction}) and a more complex breakdown using a nested transaction (Figure~\ref{nested_banking_transaction}).

Figure~\ref{flat_banking_transaction} depicts a basic banking transaction, such as a fund transfer. It outlines the sequential operations involved and the two potential outcomes: successful completion (Commit) or failure necessitating a rollback (Abort) to maintain account integrity. For instance, insufficient funds in Debt Account A during a debit would likely lead to an Abort, a scenario covered by the "Failure" path originating from Credit Account B.

Figure~\ref{nested_banking_transaction} presents a visual representation of a banking fund transfer implemented as a nested transaction, where the main operation is divided into a series of interdependent sub-operations. The overall process begins with a "Start Transaction" action, which in turn triggers the first sub-transaction, labeled "Start Sub-Transaction: Verify Funds." The purpose of this initial sub-process is to confirm the availability of adequate funds through a "CheckBalance" operation. Based on the outcome of this check, the system either proceeds to a "SufficientFunds" state, culminating in a successful "Commit Sub-Transaction: Verify Funds," or encounters "InsufficientFunds," leading to an "Abort Sub-Transaction: Verify Funds.". The subsequent step, "Start Sub-Transaction: Transfer Funds," is initiated solely upon the successful completion (commit) of the "Verify Funds" sub-transaction. This second sub-process involves the core actions of "Debit Account A" and "Credit Account B." A successful execution of both these actions results in a "Commit Sub-Transaction: Transfer Funds." Conversely, any failure during either the debit or credit operation leads to an "Abort Sub-Transaction: Transfer Funds.". The success of the entire transaction, spanning from the initial "Start Transaction" to the final "Commit Transaction," is contingent upon the successful commitment of both constituent sub-transactions. If either the "Abort Sub-Transaction: Verify Funds" or the "Abort Sub-Transaction: Transfer Funds" occurs, the entire operation is rolled back, resulting in an "Abort Transaction." This mechanism ensures the integrity of the data by reverting any partial modifications. Consequently, the final status of the primary transaction is directly determined by the combined outcomes of its sub-transactions.

We aim to perform transactions between a client and a server. The client begins a transaction that must be executed on the server. In the case of a one-phase commit, which is unidirectional, the server cannot solely decide to abort a transaction during a client request. Due to the lack of concurrency control, if the server aborts the client's request, then the client will be unaware of the server's action. The client has to make another call to know the state of the server. An improvement is to introduce a coordinator who is always alive to help serialize the actions between the client and the server. This improvement has led to the development of several atomic control protocols in SubSection~\ref{Atomic-Commit-Protocols}.

\subsection{Atomic Commit Protocols}
\label{Atomic-Commit-Protocols}

\subsubsection{ Two-Phase Commit }

Two-Phase Commit (2PC) protocol permits any party (participant) can decide to roll back its transaction (due to failure) and the entire transaction is globally reverted~\cite{coulouris2012, tanenbaum2007}. This requires a coordinator to facilitate concurrency control. The algorithm has two phases~\cite{coulouris2012}: in the first phase, each party votes to commit or abort a transaction. Once, a party has voted to commit a transaction, it is binding and immutable. After voting each party goes to the "prepared" state. In the second phase, each party in the transaction jointly executes the decision. If any party aborts or fails, then the overall transaction is aborted. The two phases ensure that all parties reach the same decision on committing or rolling back the transaction.

\begin{figure}[H] 
\centering
\includegraphics[scale=0.75]{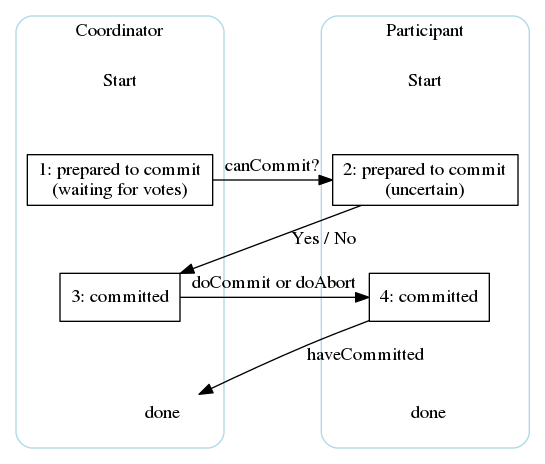}      
\caption{Two-Phase Commit~\cite{coulouris2012}. }
\label{two_phase_commit}
\end{figure}

We describe the Two-Phase Commit Protocol depicted in Figure~\ref{two_phase_commit} as follows:

\begin{itemize}
\item The coordinator makes a 'canCommit' request to every participant, and the participants respond with a vote. Only participants that send a 'yes' transition to the 'prepared' state.

\item Upon receipt of the votes from the participants, if every vote is a 'yes', then the 'doCommit' request is sent to every participant. However, if any vote is a 'no', then the current transaction on the coordinator is aborted, and then the 'doAbort' command is sent to every participant.

\item Participants, on receiving the 'doCommit', proceed to commit the transaction; otherwise, if 'doAbort' is received, then the transaction is aborted. The participant may send a 'haveCommitted' request to the coordinator to ensure the information on the coordinator can be deleted.

\end{itemize}

\subsubsection{ Three-Phase Commit }

Two-Phase Commit fails when the coordinator is dead and participants in the transaction cannot determine the state of the transaction resulting in longer polling intervals as parties try to connect and get information from the coordinator. Three-Phase Commit solves the problem by taking more rounds~\cite{coulouris2012, tanenbaum2007} as we perform a precommit round and obtain ACKS before proceeding to commit. 

\begin{figure}[H] 
\centering
\includegraphics[scale=0.75]{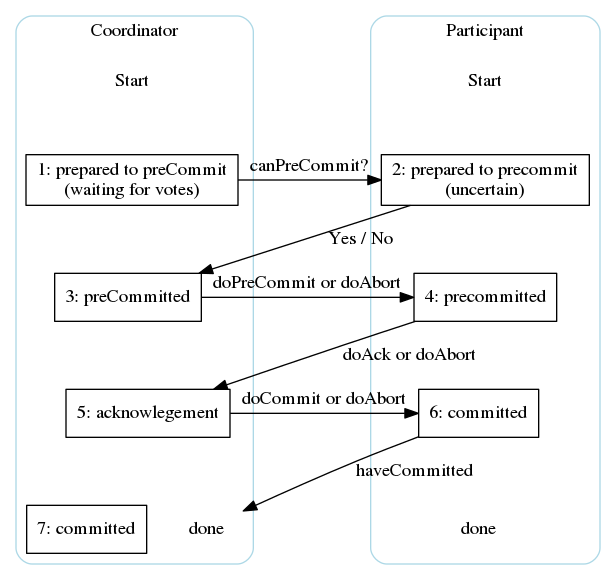}      
\caption{Three-Phase Commit~\cite{coulouris2012}. }
\label{three_phase_commit}
\end{figure}

We describe the Two-Phase Commit Protocol depicted in Figure~\ref{three_phase_commit} as follows:

\begin{itemize}
\item The coordinator makes a 'canPreCommit' request to every participant, and the participants respond with a vote. Only participants that send a 'yes' transition to the 'prepared' state.

\item Upon receipt of the votes from the participants, if every vote is a 'yes', then the 'doPreCommit' request is sent to every participant. However, if any vote is a 'no', then the current transaction on the coordinator is aborted, and then the 'doAbort' command is sent to every participant.

\item The participants, after precommitting, send an ACK to the coordinator. If there are no negative ACKs or missing ACKs, then the 'doCommit' command is sent to every participant.

\item Participants, on receiving the 'doCommit', proceed to commit the transaction; otherwise, if 'doAbort' is received, then the transaction is aborted. The participant may send a 'haveCommitted' request to the coordinator to ensure the information on the coordinator can be deleted.

\end{itemize}

\section{Routing Algorithm }
\label{routing}

For practical purposes, a thorough understanding of the following concepts may be necessary.
\begin{itemize}
\item Deep understanding of L2 / L3 networking
\item IP routing protocols such as RIP, OSPF, BGP, IS-IS, PIM.
\item Layer 2 such as 802.1d 802.lax spanning tree protocol, 802.1AB link aggregation control protocol, link layer discovery protocol, RFC 1812 IP routing.
\item Ethernet bridging and routing in Distributed software.
\item Wireless protocol (802.11, Bluetooth, Zigbee).
\item Creating a testbed (mini-internet) with the (backbone, router, switches) for testing routing protocols from scratch. Make use of dockers and VMs as shown in Network Programming in Python book.
\item Do experiments with network shaping in Docker. See \url{https://github.com/lukaszlach/docker-tc}.

\end{itemize}

\chapter{Anti-Entropy Techniques}
\label{Anti-Entropy}
\subsubsection{ \small "One of the truest tests of integrity is its blunt refusal to be compromised. -- Chinua Achebe"} 
\subsubsection{ \small "Nothing is more important than the state of the soul." -- G.K. Chesterton}

Entropy in distributed systems signifies a state of inconsistency or disagreement among the data held by different nodes. Even after a consensus algorithm has theoretically concluded and terminated, various real-world factors can lead to some nodes becoming out of synchronization with others. These contributing factors encompass the inherent complexities of distributed networks, including unpredictable network topologies, variable and often non-deterministic message delivery times, arbitrary ordering of message arrival, and the ever-present possibility of node failures. Effectively addressing these inconsistencies is paramount for maintaining the data integrity, reliability, and overall correctness of distributed systems.

The process of resolving these discrepancies can be conceptually viewed as a form of "secondary consensus" that occurs after the initial "primary consensus" protocol has aimed to establish agreement on a particular value or state. It also closely relates to the principle of eventual consistency, where updates to the system's state might not be instantaneously reflected across all nodes but are expected to propagate and converge to a consistent state over a certain period. However, in practical scenarios, this eventual convergence might take an unacceptably long duration, necessitating the implementation of active mechanisms to expedite the synchronization process and ensure timely agreement.

The robustness and overall effectiveness of a distributed software system can often be directly attributed to how well it proactively handles these inherent inconsistencies. While the specific implementation details and proprietary techniques might vary significantly between different systems, the general frameworks for reasoning about and systematically addressing entropy are based on well-established principles and algorithmic patterns.

One fundamental approach to mitigating these issues involves the deployment of background processes that periodically perform checks for discrepancies in the state held by different nodes. Upon detecting an out-of-sync condition, these processes can automatically trigger retry mechanisms to re-request missing or potentially corrupted data or employ sophisticated error correction techniques to repair incomplete or inconsistent data, thereby facilitating conflict resolution and restoring a consistent state. A straightforward yet often effective conflict resolution strategy is for a node that detects an inconsistency to request the correct or missing data from its neighboring nodes and subsequently updates its local state upon successful retrieval during these retry attempts. Well-known examples of such proactive synchronization techniques include read repair, where inconsistencies are actively detected and fixed during read operations initiated by clients, and hinted handoff, a mechanism where a temporarily available neighbor node accepts and temporarily stores writes destined for a failed node, ensuring that these writes are delivered once the intended recipient recovers and rejoins the system.

\section{Gossip Algorithm}
\label{Gossip}

The Gossip protocol~\cite{Ganesh2003, Jelasity2007, Kermarrec2007, Leito2010} presents a robust and widely applicable anti-entropy mechanism that can be effectively employed in distributed systems to promote data consistency~\cite{petrov2019}. This protocol leverages a probabilistic approach to disseminate information (messages) across a network, drawing a compelling analogy to the way rumors or information propagate through a social network. A significant advantage of the Gossip algorithm is its inherent fault tolerance and resilience to failures, even in the face of substantial and unpredictable network disruptions or node outages. While the epidemic approach, characterized by the rapid and potentially ubiquitous spread of information to all reachable nodes, is a common and intuitive implementation strategy, Gossip protocols can also be implemented using non-epidemic strategies that strategically leverage overlay network structures or maintain partial views of the system's membership to optimize message dissemination and resource utilization.

\section{CRDT}
\label{CRDT}

Conflict-free Replicated Data Types (CRDTs) represent another powerful and increasingly popular class of anti-entropy mechanisms in distributed systems. Riak, a distributed NoSQL database, was a notable pioneer in the widespread adoption and practical application of CRDTs~\cite{crdt2022, petrov2019}. A CRDT is a specialized type of data structure meticulously designed to maintain a consistent state across multiple independent replicas, regardless of the order in which concurrent operations are executed on those individual replicas. This inherent order-independence property makes CRDTs exceptionally well-suited for scenarios involving collaborative applications and seamless remote synchronization (replication) of data across numerous geographically distributed devices, a task that can be inherently complex to manage for data consistency using traditional concurrency control mechanisms. Mathematically, CRDTs can be formally characterized as partially ordered monoids that possess a well-defined lattice structure. To ensure their conflict-free nature, they must satisfy the algebraic properties of commutativity (the order in which operations are applied does not affect the final state) and idempotence (applying the same operation multiple times has the same effect as applying it only once).

CRDTs offer several compelling advantages for building highly available and eventually consistent distributed applications:

\begin{itemize}
\item \textbf{Eventual consistency:} They provide a strong guarantee that all replicas of the CRDT will eventually converge to the same consistent state once all locally generated operations have been successfully propagated and applied to all other replicas in the system.
\item \textbf{Preserve ordering of the data:} While the order of concurrent *operations* might not be critical for achieving eventual consistency, certain types of CRDTs are specifically designed to inherently preserve the logical or intended ordering of the *data* they are managing, which is crucial for applications requiring ordered data structures.
\item \textbf{Local-first application:} Operations performed on a CRDT can be applied locally on a replica immediately without the need to wait for synchronization with other replicas. This "local-first" characteristic significantly improves the responsiveness and perceived performance of applications, leading to a better user experience, especially in high-latency network environments.
\end{itemize}

A diverse range of CRDTs has been developed, each tailored to manage different types of data and support specific sets of operations. Common examples include map CRDTs (such as the Last-Write-Wins Map - LWW-Map), set CRDTs (such as the Grow-Only Set - G-Set), and many other specialized data structures designed for various consistency and concurrency control requirements~\cite{crdt2022, petrov2019}.

We have implemented a version of LWW-Map is shown in \href{https://github.com/kenluck2001/DistributedSystemReseach/blob/master/textbook/lamport1-majority-voting8.c}{lamport1-majority-voting8.c}. Basically, we make updates and use the largest timestamp of processes to determine the final value of a map. 

The provided C source code implements a Grow-only Counter (GCounter) as shown in \href{https://github.com/kenluck2001/DistributedSystemReseach/blob/master/textbook/crdt-gcounter.c}{crdt-gcounter.c}, a type of Conflict-free Replicated Data Type (CRDT), utilizing the Message Passing Interface (MPI) for parallel operation. Each parallel process maintains a local state stored in the `GCounter` structure, where the `counts` array tracks the contributions of every process on a per-index level. The core CRDT operation is the merge, which is performed using the collective primitive `MPI\_Allreduce` with the reduction operator `MPI\_MAX`. This operation ensures that every process receives the global state where each element of the merged array is the maximum of the corresponding local element from all contributing arrays, correctly satisfying the GCounter's merge property: $merge(A, B) = max(A_i, B_i)$. Finally, the `gcounter\_value` function calculates the counter's total by performing a summation over all indices of the merged array; while the true CRDT merge logic uses `MPI\_MAX`, the summation provides a shorthand method to check that the merged arrays have similar content and allows for a final, consistent global value check across all processes for demonstration purposes.

\href{https://github.com/kenluck2001/DistributedSystemReseach/blob/master/textbook/lww-map.c}{lww-map.c} implements a Last-Write-Wins Map (LWW-Map), a type of Conflict-free Replicated Data Type (CRDT), using the Message Passing Interface (MPI) to handle state replication and global merging across multiple parallel processes. The basis of the code is to allow independent, concurrent updates (puts and deletes) to a key-value store on different nodes, with a deterministic merge function that resolves conflicts using a logical timestamp and a node ID tie-breaker. The core data structure, `LWWMap`, holds an array of `Entry` structs, where each entry records the key, value, a logical timestamp (`ts`), the node ID (`node`) that made the update, and a tombstone flag (`tomb`) to mark deletions. The `put` and `del` functions increment the map's local clock $(m \rightarrow clock)$  before writing the entry, ensuring newer operations have a higher timestamp. The critical functionality is handled by the `global\_merge` function: Rank 0 first collects the local state arrays from all other ranks using `MPI\_Gather` to determine the size and subsequent `MPI\_Recv` (or local `memcpy` for itself) to gather all individual entries into a single large buffer. Rank 0 then performs the actual merge logic: it iterates through the collected entries and for each unique key, it compares the timestamp and node ID of the new entry against the currently merged entry using the `newer` function, keeping only the one with the later timestamp or higher node ID. Finally, Rank 0 broadcasts the resulting, fully merged `LWWMap` state back to all other processes using `MPI\_Bcast`, ensuring every node is synchronized. The `main` function demonstrates a conflict scenario where different ranks update or delete the same keys ('a', 'b', 'c'), followed by a global merge and a call to `print\_visible`, which displays the final, consistent state across all ranks, omitting any entries marked by a tombstone.

\section{Operational Transformation}
\label{OpTransform}

Operational Transformation (OT) is a sophisticated technique primarily employed in collaborative editing systems, such as shared document editors, to maintain consistency across multiple concurrent users editing the same document simultaneously. Similar to CRDTs, OT aims to resolve conflicts that arise from concurrent operations performed by different users by transforming the semantics of these operations based on the context of previously executed operations. However, OT typically deals with ordered and sequential operations on linear data structures like text or rich text documents, and its underlying mathematical principles and implementation complexities differ significantly from those of CRDTs, which are generally more focused on unordered sets or map-like data structures.

TODO: PROVIDE IMPLEMENTATIONS

\section{Ancillary Structures}
\label{ancillary}

Ancillary data structures, while not directly mechanisms for achieving consensus or propagating updates, can play a vital role in efficiently verifying whether the state held by different nodes in a distributed system is in disagreement or has become inconsistent. Structures like the Merkle tree provide an efficient method for identifying and potentially facilitating the resolution of such conflicts. A Merkle tree is a tree-like data structure where each non-leaf node is a cryptographic hash of its child nodes, and the leaf nodes are cryptographic hashes of individual data chunks or blocks. This hierarchical hashing mechanism allows for the efficient and secure verification of data integrity and consistency across distributed systems. We can effectively identify disagreements among nodes if they fail to produce matching results during audit or consistency proofs based on their respective Merkle trees. Let us delve into the details of these proofs:

\begin{itemize}
\item \textbf{Audit proof:} An audit-proof enables a node to efficiently and cryptographically verify that a specific, single data chunk exists within a Merkle tree held by another node, without the necessity of downloading the entire potentially large dataset. This is achieved by the requesting node providing the path of hashes from the leaf node corresponding to the data chunk up to the root of the Merkle tree, allowing the verifying node to recompute the root hash and compare it with its own.
\item \textbf{Consistency proof:} A consistency proof allows a node that holds one version of a Merkle tree (representing the state of the data at a particular point in time) to efficiently and cryptographically verify that another node holding a potentially different version of the same Merkle tree represents a consistent historical version of its own tree. This enables efficient comparison of different states of the same dataset over time, identifying points of divergence or ensuring that one version is indeed an ancestor of another.
\end{itemize}

Consistency verification using Merkle trees significantly facilitates:

\begin{itemize}
\item \textbf{Data verification:} Ensuring the integrity and cryptographic correctness of the data held by individual nodes by verifying the consistency of their Merkle trees or specific branches thereof.
\item \textbf{Data synchronization:} Efficiently identifying the specific data chunks or sub-trees that differ between nodes by comparing their Merkle tree structures, allowing for targeted synchronization of only the differing data and minimizing the overhead of transferring large, identical datasets.
\end{itemize}

My understanding of how the Merkle tree functions is that it provides a highly efficient and cryptographically sound way to pinpoint the precise data chunks or segments that exhibit differences between nodes in a distributed system. A common initial step is for nodes to exchange the root hash of their respective Merkle trees as lightweight metadata. If the root hashes match, there is a very high probability (depending on the cryptographic hash function used) that the entire underlying dataset is consistent across the nodes. However, if the root hashes differ, it definitively indicates a mismatch in the data. Given the hierarchical structure of the Merkle tree, it allows for a more granular and efficient comparison process. Nodes can recursively traverse down the tree, comparing the hashes of child nodes at each level. When a hash mismatch is detected at a particular level, it indicates that the inconsistency lies within the subtree rooted at that node. This allows nodes to selectively request and synchronize only the specific sub-items or data chunks that are inconsistent, significantly reducing the network bandwidth and processing overhead compared to transferring and comparing entire datasets.

The intriguing and often debated aspect is the claim that Merkle trees can, in certain specific scenarios, facilitate conflict resolution directly, going beyond simply identifying discrepancies and triggering subsequent retry or synchronization mechanisms. While my intuition aligns with the understanding that conflict resolution typically necessitates a separate, higher-level mechanism to decide which version of conflicting data to ultimately retain or merge, there might indeed exist specific scientific papers or specialized techniques that leverage the inherent structural properties or metadata embedded within Merkle trees to automate certain limited types of conflict resolution, particularly in specific data models or application contexts. I would be very interested in learning about such scientific papers that rigorously detail methods for automatic conflict resolution directly derived from the structure or properties of Merkle trees, rather than just using them for efficient detection of inconsistencies.

The fundamental challenge of resolving conflicts in distributed systems often involves navigating the inherent trade-off between achieving strong consistency and maintaining high availability.

\begin{itemize}
\item \textbf{Strong consistency:} This stringent approach ensures that every node in the distributed system is effectively locked or prevented from serving potentially stale data until all nodes have successfully acknowledged and applied an update to a new value. This guarantees that all read operations will reflect the most recent write operation, providing a linearizable view of the data. However, achieving strong consistency can significantly impact the system's availability and performance, especially in the presence of network partitions or transient node failures, as the system might become unavailable if a quorum of nodes cannot be reached.
\item \textbf{Eventual consistency:} This more relaxed consistency model represents a deliberate trade-off that prioritizes high availability and partition tolerance over immediate consistency. In eventually consistent systems, updates are applied to some replicas, and the system provides a probabilistic guarantee that all replicas will eventually converge to the same consistent state at some point in the future, assuming no further updates occur. During the period before complete convergence, different replicas might temporarily serve different (potentially stale or outdated) data. A common and often desirable consistency guarantee within eventually consistent systems is "read-your-writes" consistency, which ensures that a client will always see the results of its own immediately preceding write operations, even if other replicas in the system have not yet fully processed the update.
\end{itemize}

\section{Error Correction Code}
\label{error-correction}

Failures are an inherent and unavoidable aspect of operating large-scale distributed systems, and the field of coding theory provides a rich and well-established set of mathematical schemes and algorithms for detecting and recovering from these errors, thereby ensuring data integrity, reliability, and continuous availability. We will focus on two prominent and widely used schemes for error recovery in the following subsections: Erasure Coding (Subsection~\ref{erasure-coding}) and Multiple Description Coding (Subsection~\ref{multiple-description-coding}).

For readers interested in a more in-depth exploration of the theoretical underpinnings and practical applications of various error recovery codes, the following highly regarded texts are recommended as valuable resources:

\begin{itemize}
\item Error Correction Coding: Mathematical Methods and Algorithms by Todd K. Moon
\item Fundamentals of Error-Correcting Codes by W. Cary Huffman and Vera Pless
\item Turbo Code Applications: A Journey from a Paper to Realization by Keattisak Sripimanwat
\end{itemize}

\subsection{Erasure Coding}
\label{erasure-coding}

Erasure coding (EC) is a sophisticated and highly effective redundancy scheme specifically designed to enhance data reliability and provide robust fault tolerance, particularly in distributed storage systems where multiple storage drives or nodes can fail unexpectedly and concurrently. This technique offers a significantly more storage-efficient alternative to traditional full replication strategies, such as those used in conventional RAID (Redundant Array of Independent Disks) configurations, by allowing for a configurable and often substantially lower storage overhead while still providing comparable or even superior levels of data durability and availability.

The fundamental principle behind erasure coding involves systematically organizing data into a set of data segments or blocks and then generating a corresponding set of redundant parity blocks. These parity blocks contain carefully encoded complementary information derived from the original data blocks, which can be mathematically leveraged to reconstruct the original data in the event of the loss or corruption of some of the data blocks or parity blocks themselves. An Erasure Coding algorithm intelligently utilizes the remaining intact data blocks and the parity blocks to both detect the occurrence of failures (erasures) and, crucially, reconstruct the lost data, ensuring data integrity and availability. It is generally recognized as impractical and highly wasteful of storage capacity to maintain a dedicated parity block for each original data block. Instead, a common and more efficient approach is to group a set of $n$ original data blocks together and generate a smaller set of $m$ parity blocks that collectively provide the necessary redundancy for error recovery. For such an $(n, m)$ erasure coding configuration to function effectively in a real-world storage system, a total of $n + m$ storage drives or nodes are required. The key advantage is that the storage overhead is determined by the ratio of $m$ to $n$, which can be significantly lower than the $1:1$ overhead of full replication while still offering substantial fault tolerance (the ability to recover from the loss of any $m$ drives).

\subsection{Multiple Description Coding}
\label{multiple-description-coding}

Multiple Description Coding (MDC) represents another important class of redundancy schemes specifically designed to enhance resilience to failures and mitigate the impact of network issues by strategically splitting a single source data stream into multiple, inherently redundant substreams or descriptions. These individual segments or descriptions might be encoded with varying levels of detail, and resolution, or possess distinct characteristics tailored for different transmission or reception conditions, but they all contain a similar core of essential information about the original data. The receiver (callee) then employs a variety of techniques, often dynamically adapting based on real-time quality of service (QoS) metrics, to selectively utilize the available received segments for reconstructing the original data stream.

A prominent and widely deployed real-world example of Multiple Description Coding principles in action is adaptive bit rate (ABR) streaming for video content delivery over the internet. In ABR, a single video is encoded into multiple independent streams with varying levels of quality and resolution (e.g., low, medium, high definition). The video player (callee) continuously monitors prevailing network conditions, such as the available internet speed, network bandwidth, and the level of contention on the network path. Based on these dynamically measured QoS metrics, the player intelligently and seamlessly switches between requesting and decoding segments from the different resolution streams. This adaptive approach allows the player to provide the best possible viewing experience under the current network conditions, gracefully degrading the video quality if the network deteriorates and automatically improving it when the network capacity allows.

\footnote{ \footnotesize See an illustrative example of optimizing a distributed data store for operation at significant scale in the \href{https://www.uber.com/en-CA/blog/how-uber-optimized-cassandra-operations-at-scale/}{Uber blog}}

\chapter{Peer-to-Peer Computing}
\label{Peer-to-Peer-Conn}
\subsubsection{ \small "In the midst of chaos, there is also opportunity." -- Sun Tzu } 
\subsubsection{ \small "Life is a wheel of fortune and it's my turn to spin it." -- Tupac Amaru Shakur}

In the peer-to-peer (P2P) paradigm, each network node functions as both a client and a server simultaneously. This allows any node to communicate directly with its neighbors without the necessity of a dedicated central server. This decentralized arrangement inherently avoids a single point of failure, enhancing the system's robustness. However, some hybrid peer-to-peer systems exist where a central server is utilized for initial orchestration and connection establishment. These servers store information about the peers present in the network, facilitating the discovery of other nodes.

In contrast, the traditional client-server paradigm restricts direct communication between clients. For one client to interact with another, it typically needs to update the state on the central server, which then allows other clients to connect to the server and retrieve the shared information. In this model, the server becomes a critical component for all communication, making it a significant single point of failure for the entire system.

Peer-to-peer computing offers notable benefits, including reduced storage and computational costs, achieved by distributing the burden of computation and data storage across several participating nodes. The applications of peer-to-peer computing span a wide range of domains, including distributed file systems (e.g., OceanStore~\cite{Kubiatowicz2000}, PAST~\cite{Druschel2001}, CFS~\cite{DabekKKMS2001}), file sharing networks (e.g., BitTorrent~\cite{BitTorrent2019}, FastTrack~\cite{FastTrack2022}), and large-scale Grid computing platforms~\cite{GridComputing2022}.

\begin{figure}[H]
\centering
\includegraphics[scale=1.2]{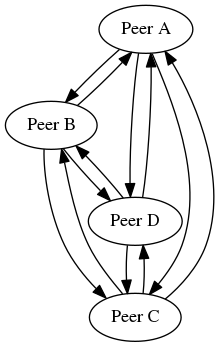}
\caption{Network of nodes in a peer-to-peer network.}
\label{p2p_architecture}
\end{figure}

We present a peer-to-peer network of 4 nodes (peers) named "A", "B", "C", and "D" as shown in Figure~\ref{p2p_architecture}. Each node can connect with any other node in the network. There is no central server or intermediary required for one peer to interact with another node resulting in decentralized network.

@TODO: IMPLEMENTATION

We will now discuss the categorization of peer-to-peer systems to address core concepts such as overlay networks, unstructured P2P systems, and structured P2P systems, as covered in Sections \ref{Resilient-Overlay-Networks}, \ref{Unstructured-P2P-Systems }, and \ref{Structured-P2P-Systems } respectively.

\section{Resilient Overlay Networks}
\label{Resilient-Overlay-Networks}

An overlay network provides a dynamic, just-in-time network infrastructure built on top of an existing network layer (like the Internet) to facilitate decentralized communication in a scalable manner. It achieves this by implementing its own routing and packet transfer functionalities at the application level, abstracting away the complexities of the underlying physical network~\cite{caesar2006, david2001}.

Overlay networks must be designed to handle a high rate of client disconnections, which can occur due to various factors such as poor network conditions, limited available bandwidth, and the need for dynamic balancing of processing and bandwidth capacity across the participating nodes~\cite{castro2005}. Consequently, there is a significant need for the development of resilient overlay networks that enhance the reliability of the network and make it robust to node and link failures~\cite{david2001}. Nodes within resilient overlay networks often incorporate self-healing capabilities. Such nodes can perform quality of service (QoS) checks on the communication links they use and dynamically re-route packets as necessary to circumvent failures and reduce latency.

A key characteristic of overlay networks is that the participating nodes do not need to reside within the same local network. They can be geographically dispersed across various administrative domains and networks. Due to the dynamic redirection of packets on the fly, the maintenance of a routing table at the overlay level becomes essential.

When communicating across different Internet Service Providers (ISPs), one might naively assume that Border Gateway Protocol (BGP) routing records are consistently up-to-date and reflect the optimal paths. However, this is often not the case, as the entries in an ISP's routing table can be influenced by a multitude of factors, including financial agreements between ISPs, the time it takes for BGP to converge after a routing update, and even political considerations~\cite{david2001}.

The routing table maintained at the application level by the overlay network can potentially identify and utilize more efficient and reliable communication links compared to the underlying BGP routes. Content Delivery Networks (CDNs) are a prominent example of how overlay networks are leveraged to improve offsite caching and the efficient delivery of web resources to end-users~\cite{david2001}.

\section{Unstructured P2P Systems}
\label{Unstructured-P2P-Systems }

Unstructured overlay networks are characterized by the absence of a predefined, deterministic structure for organizing nodes and data. Consequently, they do not inherently provide efficient query-matching mechanisms for locating specific data within the network~\cite{castro2005}. Examples of P2P systems that utilize unstructured overlays with a random graph topology include Gnutella~\cite{taylor2009}, the now-defunct Kazaa, and Freenet~\cite{clarke2001}. In such networks, a typical search query is flooded across the network to a subset or all neighboring nodes. Each node receiving the query then compares it against the metadata or content it holds.

\section{Structured P2P Systems}
\label{Structured-P2P-Systems }

In contrast to unstructured P2P systems, structured overlay networks impose a specific, well-defined structure on the network topology to facilitate efficient query matching and data discovery~\cite{castro2005}. This efficient search capability is achieved by enforcing constraints on how the network is constructed and how data is indexed and located. In a structured P2P system, each node is typically associated with a unique key or identifier, and the actual data or metadata is stored at the node whose key is closest to the key of the data item. Searching for a specific data item involves querying the network using its associated key. The network's underlying structure, often based on sophisticated data structures like Distributed Hash Tables (DHTs), ensures that queries can be routed efficiently to the node(s) likely to hold the desired information. Prominent examples of P2P systems that leverage structured overlay networks include Chord~\cite{stoica2003}, Tapestry~\cite{zhao2004}, Kademlia~\cite{kademlia}, and Pastry~\cite{rowstron2001}.


\chapter{Practical Formal Verification of Distributed Systems}
\label{Formal-Verification}
\subsubsection{ \small "It is better to be roughly right than precisely wrong." -- John Maynard Keynes }

Distributed systems present significant testing challenges due to the multitude of realistic and abnormal scenarios they must handle. This complexity is amplified by the inherent mechanisms within distributed systems designed to detect various types of failures, which can range from transient issues to more persistent Byzantine faults. Effectively addressing these failures is a primary objective in designing robust distributed systems that prioritize fault tolerance as a core principle. However, the implementation of mechanisms to mitigate these failures often introduces increased complexity into the system's architecture. This added complexity can inadvertently conceal subtle bugs, which can have severe consequences in mission-critical applications. Formal verification offers a powerful approach to verify, at the design level, that these potential problems do not exist. Careful attention must be paid to ensure that there is no discrepancy between the system's implementation and its formal specification. Once a system has undergone and successfully passed formal verification, a higher degree of confidence can be placed in the absence of critical bugs within the system.

While formal verification offers strong guarantees, it can be an expensive undertaking, primarily due to the significant effort required to develop a precise formal specification and to rigorously prove that the software implementation adheres to this specification. However, a substantial number of bugs can be effectively caught through less formal but still valuable testing methodologies such as unit tests, integration tests, and fault-injection tests. These testing approaches can help to uncover subtle bugs and significantly improve the overall quality of the distributed software. Software testing, in general, is less formal compared to the rigorous mathematical proofs employed in formal verification. Unit tests should be strategically designed to identify potential failure modes within individual components and to achieve high coverage of the problem space related to these failures. Integration tests play a crucial role in verifying the behavior of the system as different nodes communicate with each other over a simulated or real network channel within a cluster. A particularly insightful type of integration test is the fault-injection test, where engineers intentionally introduce specific failure scenarios into the running system and carefully observe how the collaborating components behave under these adverse conditions. For instance, in a distributed data store, a fault-injection test might involve simulating a node failure to observe if there is any data loss or corruption.~\href{https://kenluck2001.github.io/blog_post/metamorphic_testing_in_a_nutshell.html}{Metamorphic testing} is another valuable technique that can enhance test coverage by leveraging known relationships between different inputs and outputs and checking for violations of invariants that should hold even in the presence of failures. Several notable examples of robust testing practices in the industry include~\href{https://netflixtechblog.com/the-netflix-simian-army-16e57fbab116?gi=c94d70504e80}{Netflix Simian Army}, a suite of tools designed to randomly introduce failures into Netflix's cloud infrastructure to ensure resilience;~\href{https://jepsen.io/](https://jepsen.io/}{Jepsen}, a rigorous testing framework for distributed databases that focuses on uncovering consistency violations under various fault conditions; and~\href{https://verdi.uwplse.org/}{Verdi}, a research project focused on the formal verification of distributed systems implementations.

Formal verification serves as an overarching term for a collection of techniques aimed at mathematically proving whether a system's design or implementation adheres to a precisely defined set of specifications. A specification is a written document that describes the intended behavior and properties of a system in an unambiguous manner. While human language is often verbose and can include more information than is strictly necessary to convey the core requirements, mathematics offers a more concise and precise language for describing most phenomena in the physical and computational worlds. Unfortunately, even mathematical proofs can sometimes rely on informal arguments expressed in human language. Mathematical logic, particularly propositional logic, and temporal logic, can help to eliminate this redundancy and imprecision, leading to more rigorous and unambiguous specifications and proofs, which form the basis of formal methods.

The primary goal of formal verification is to identify potential errors and flaws in the system's design early in the development lifecycle, based on the formal specifications. Therefore, the specification should be carefully derived from the core requirements of the system to be truly useful in detecting critical issues. The specification must be kept as simple and focused as possible, so that the constraints on the system's behavior can be clearly and precisely defined, thereby mitigating the risk of the "second-system effect" (where an over-engineered second attempt at a system introduces more problems than it solves). A useful rule of thumb is to exclude any information from the specification that does not directly contribute to understanding the fundamental behavior and properties of the system, as this extraneous information is less likely to reveal critical violations. Creating a precise and comprehensive specification is a significant challenge in software development. Careful consideration must be given to composing a realistic specification that accurately reflects the system's requirements and subsequently writing source code that provably satisfies this specification. Ideally, the effort involved in writing the formal specification should be relatively less than the effort required to write the actual source code that implements and verifies the specified behavior, to avoid doubling the programmer's workload. Temporal logic is a specific type of formal logic that is particularly well-suited for proving the liveness properties of concurrent and distributed systems, ensuring that the system will eventually reach desirable states.

\section{Model-Level Verification}

One prominent model-based formal verification scheme specifically designed for distributed systems was developed by the renowned computer scientist Leslie Lamport. This scheme is called the Temporal Logic of Actions (TLA+). TLA+ allows for the description of the system under scrutiny using a set of mathematical equations and logical predicates that capture its state transitions and behavior over time. Notably, TLA+ also possesses non-temporal (state-based) semantics, making it a versatile formalism suitable for specifying a wide range of distributed systems, from communication protocols to concurrency control algorithms. Several tools, such as the TLA+ Toolbox and the interactive proof assistant Coq, support the specification and verification of systems using TLA+.

\section{Code-Level Verification}

Code-level verification takes a different approach compared to traditional model-level formal verification, where an abstract model of the problem space is created and checked for violations of safety and liveness properties. Instead, code-level verification directly examines the source code of the system, often involving running MPI/C source code within a specialized verification scheduler. A significant challenge in verifying concurrent programs is the potentially exponential number of possible interleavings of the execution of parallel threads or processes. To address this, code-level verification techniques often employ partial order reduction (POR) algorithms to identify and eliminate redundant or equivalent interleavings, thereby significantly reducing the state space that needs to be explored~\cite{Vo2009}.

A notable tool in the domain of code-level verification for distributed systems is the ISP Formal Verification Tool (though the provided text does not give a full name or link, "ISP" likely refers to a specific research project or tool). These types of tools often focus on detecting critical concurrency bugs such as deadlocks and runtime error violations directly within the source code.

Code-level verification aims to bridge the "gulf of execution" that can exist between abstract model checking and the behavior of actual running source code by directly targeting the implementation. This alignment increases confidence that the verification results accurately reflect the real system's behavior. The primary focus is often on analyzing the interleavings of parallel threads to uncover subtle concurrency-related bugs.

While code-level verification offers the advantage of directly analyzing the implementation, model-based verification can provide productivity gains by allowing engineers to create abstract models of their concurrent designs and verify their properties before writing the full production code. However, code-based verifiers may face limitations in handling the full expressiveness of temporal logic specifications compared to dedicated model checkers.

One such tool is \href{https://valgrind.org/}{Valgrind}. We show how to use 3 tools associated with Valgrind. The following description is done with the assumption that the user has previously installed Valgrind on the system.

\subsubsection{memcheck} 
We can analyze the memory usage and health of a any program say test5.c in our example, which utilizes the OpenMPI library. This assumes that Valgrind is already installed on the system.

\begin{verbatim}
mpicc -g test5.c -o test5
mpirun -np 4 valgrind --tool=memcheck --leak-check=full ./test5
\end{verbatim}

The output is given as follows.
\begin{verbatim}
 Leak Summary (Printed at the End of the Program):

==<PID>== HEAP SUMMARY:
==<PID>==     in use at exit: <bytes> bytes in <blocks> blocks
==<PID>==   total heap usage: <allocs> allocs, <frees> frees, 
<bytes> bytes allocated
==<PID>==
==<PID>== LEAK SUMMARY:
==<PID>==    definitely lost: <bytes> bytes in <blocks> blocks
==<PID>==  indirectly lost: <bytes> bytes in <blocks> blocks
==<PID>==    possibly lost: <bytes> bytes in <blocks> blocks
==<PID>==      still reachable: <bytes> bytes in <blocks> blocks
==<PID>==         suppressed: <bytes> bytes in <blocks> blocks
\end{verbatim}

HEAP SUMMARY: Provides an overview of heap memory usage.
\begin{enumerate}
\item in use at exit: The amount of heap memory that was still allocated when the program terminated. A non-zero value suggests memory leaks.
\item total heap usage: The total number of allocation and deallocation operations and the total bytes allocated.
\end{enumerate}

LEAK SUMMARY: Categorizes unfreed memory blocks:
\begin{enumerate}
\item definitely lost: Memory that your program allocated and then lost all pointers to. This is a clear memory leak that you should fix.
\item indirectly lost: Memory that is pointed to only by definitely lost memory. Fixing the "definitely lost" blocks will usually also resolve these.   
\item possibly lost: Memory that might still be reachable through some complex pointer paths, but Valgrind isn't entirely sure. Investigate these, as they often indicate leaks.  
\item still reachable: Memory that was allocated but never freed, but the program still holds pointers to it at exit. While technically not a leak in the strictest sense, it's often good practice to free this memory as well. You can suppress these reports with --show-reachable=no.  
\item  suppressed: Leaks that were ignored due to entries in suppression files.  
\end{enumerate}

It is better to reduce the size and number of blocks that are 'definitely lost or indirectly lost' as part of improving your source code.

\subsubsection{Helgrind and DRD}

Helgrind and DRD are primarily designed to analyze programs using POSIX threads (pthreads) for concurrency errors and violations. It can provide a sanity check for the OpenMPI program, but it was not originally designed for the message-passing paradigm of our underlying runtime.

\begin{verbatim}
mpicc -g test5.c -o test5
mpirun -np 4 valgrind --tool=memcheck --leak-check=full ./test5
\end{verbatim}

To get better output, get separate logs from each running process using the command provided below.

\begin{verbatim}
mpirun -np 4 valgrind --tool=helgrind --log-file=helgrind.%p.log ./test5
mpirun -np 4 valgrind --tool=drd --log-file=drd.%p.log ./test5
\end{verbatim}

Interpreting the Output from Helgrind and DRD:
\begin{enumerate}
\item Valgrind (Helgrind or DRD) will output any detected threading errors or potential deadlocks to the standard error (stderr).
\item Carefully examine the output. Look for reports of lock contention, potential deadlocks (in Helgrind), or data races (in DRD).
\item Remember that these tools are interpreting the MPI processes as independent programs that might be using internal threading within the MPI library itself. The deadlocks they detect might be related to these internal threads or might be misinterpretations of MPI communication.
\end{enumerate}

\chapter{Miscellaneous}
\label{Miscellaneous}
\subsubsection{\small "Rarely affirm, seldom deny, always distinguish." -- Thomas Aquinas } 
\subsubsection{ \small "Any subject can be made interesting, and therefore any subject can be made boring." -- Hilaire Belloc}

\section{Accelerated Computing}
\label{accelerated-computing}

There are several computing architectures where specialized processors are designed to relieve the burden of computationally intensive tasks. Custom accelerators are crafted to offload computationally expensive tasks in an optimized manner onto dedicated hardware. Accelerators are experiencing increasing adoption. For example, the Graphics Processing Unit (GPU) is optimized for rapid matrix-based operations and can function as a co-processor alongside the system's main processor. Furthermore, specialized cryptographic accelerators exist, such as the \href{https://www.intel.com/content/www/us/en/developer/articles/technical/advanced-encryption-standard-instructions-aes-ni.html}{Intel Advanced Encryption Standard Instructions (AES-NI)}, which are utilized by cryptographic libraries like \href{https://www.openssl.org/}{OpenSSL} and \href{https://www.libressl.org/}{LibreSSL}. Offloading specific tasks to these accelerators can be viewed as a form of distributed system operating between the main processor and these specialized hardware components.

\section{Storage Systems}

The Unix file system provides a convenient hierarchical structure for the storage of data in files and directories. It was not built by default to be fault-tolerant. However, using a RAID structure, where many disks work together as a replicated data store, can help to achieve resilience in the face of failure. There is also a problem related to the increased storage cost, where the metadata per directory becomes significant at petabyte scales and beyond. There is room for optimization, as the metadata has to be loaded into memory from the disk to access the file. Due to the locality of data, this is a significant bottleneck for scalability.

Every volume, despite its sophisticated architecture, includes the following core features: a data file (the actual stored data), an index file (a searchable data structure), a journal file (which allows for persistence between restarts), a checkpoint (a snapshot that provides safe points for recovery after failures), and a client (an application that has access to the file system). These features have various names across different file systems, such as HDFS~\cite{shvachko2010hadoop}, Google File System~\cite{ghemawat2003google}, and Tectonic~\cite{PanTectonic2021}. Some file systems~\cite{rosenblum1992} allow for sequential modification to a disk in a log-like manner. Another way to categorize storage systems is as single-tenant (Haystack~\cite{Beaver2010}, F4~\cite{Muralidharf42014}, HDFS~\cite{shvachko2010hadoop}) or multi-tenant (Tectonic~\cite{PanTectonic2021}). A number of these storage systems utilize variants of Paxos and Raft to achieve distributed consensus across volumes. For more information on distributed consensus algorithms, please refer to Chapter \ref{Dist-Con-Algo}, Sections \ref{paxos} and \ref{raft}.

There is room for optimizations to improve storage, leading to the development of some distributed storage systems. Haystack~\cite{Beaver2010} improves upon the Unix file system by saving metadata in memory, thereby making lookups faster, while reading the actual file saved in the volume involves disk operations, thereby increasing throughput. The metadata is spread over a wide area of memory, allowing for compact representation in the Haystack store, rather than each directory storing the metadata. However, it is necessary to ensure that there is sufficient memory to load a significant amount of the metadata for lookup purposes. When an object is saved in a volume in a Haystack store, it is replicated to many secondary volumes for redundancy. The Haystack directory maps each volume to the metadata to facilitate reading the actual file using the metadata. The Haystack cache is used to increase the lookup speed of popular objects. This is ideal for immutable blobs, hence it follows the philosophy of write once, read always, and rarely update.

Haystack is deficient in utilizing the access patterns of requests for objects, and even the Haystack cache is not sufficient for granular control of access patterns. Hence, F4~\cite{Muralidharf42014} was developed to allow for incorporating access patterns into the design of a store. Rather than a RAID-like replication scheme, F4 utilizes distributed erasure coding and Reed-Solomon codes to ensure smarter, lower-cost replication to several secondary volumes. Objects saved in the store begin as "hot" and then "warm" over time. The idea is based on the pattern that newer objects are more likely to be accessed, while older objects are less likely to be accessed. This concept is known as temperature zones. Volumes are structured in a way that once a memory limit for a volume is reached, it is closed for writing and becomes read-only. These volumes are organized into temperature zones to take advantage of access patterns. There are also many systems (Pelican~\cite{Balakrishnan2014}, RADOS~\cite{weil2007}) that allow for intelligent use of access patterns, and a system (CRUSH~\cite{weil2006}) that provides the option to specify weights for access patterns. Tectonic~\cite{PanTectonic2021} is an evolution of the Haystack storage system, as it introduced multi-tenant features. Each tenant has its own namespace and delegates metadata lookup to a dedicated key store.

\section{Coding Philosophy}
\begin{itemize}
\item Made use of mpi, so we don't have to consider the low-level socket networking stack and their quicks, between IPV4, and IPV6.

\item Our implementation of Distributed systems creates a set of processes for the client. This may not be standard in the Paxos algorithm. We abstract it in our source code for ease of use.

\item Our philosophy has been to think locally and act globally. We do computations on the node and communicate with other nodes by messaging.

\item Production-grade Distributed systems should follow the best Software engineering practices. We are not optimizing our source code for production, but pedagogy.

\item Rather than communicating by sharing memory, it is better to share memory by communicating.

\item For the Paxos algorithm when using Unix timestamps as the round number or ballots for their monotonically increasing properties, then a necessary prerequisite is to synchronize the time settings on at least the set of proposers, or across the cluster, but that may be unnecessary. This paradigm was heavily used in our implementations.

\item Organizing the processes into groups with custom communicators. This allows for targeted synchronization for grouped processes without impacting the total processes in the application. This logic allows for precise control of a group of processes.

\item When trying to create an array of atomic counters. It is desirable to utilize an array of shared pointers, rather than an array of shared values.

\item We can cancel pending requests and tune the criteria for quorum based on business needs. This would impact how resilience of the Distributed Systems.

\item We use pooling on receiving the message and checking each tag, rather than waiting on specific tags to make code modular. With my quest to go low level, rather than use C++, I tried objected-oriented design in C to enhance modularity but abandoned the idea due to loss of type safety.

\item Always pool on waiting reads in a busy-wait style.

\item Make use of a simple structure. Even our log for sequence Paxos is not a log, but an abused linked list with some atomic primitives.

\item Our sequence Paxos uses a single Paxos on each item that the proposer will send. Unfortunately, our logic is restricting to only the possibility of having one proposer.

\item Retrieving messages and probing to check the tag of messages to identify a specific event. Busy waiting is used to retrieve messages on an irecv. Otherwise, only the last sent is received on polling. This can be a bug where you retrieve the same message multiple times.

\item It is good to take steps to avoid both deadlocks and livelocks. Deadlock can happen in a mismatched message order between send and receive especially in blocking mode. It is possible in non-blocking mode to consider how request objects are owned between the receiving and sending nodes.

\item Livelock is possible too in a non-blocking case when we pull in a busy wait manner. We exit from the end of the loop when we have received the expected number of messages. It can be sensible to keep track of the number of exchanged messages to force an exit from the endless loop.

\item There are problems with passing pointers across nodes. This is because we don't have a universal shared memory. Always keep pointers local as a lack of distributed memory makes indirection on a pointer useless.

\item Use timeouts to prevent resources from waiting indefinitely for computing.
\end{itemize}

\subsection{Review of Selected Source Codes}
\begin{itemize}

\item \href{https://github.com/kenluck2001/DistributedSystemReseach/blob/master/textbook/leader-election3.c}{leader-election3.c}
This is a simplified version of our adaptation of the bully algorithm where the first candidate becomes the node to call an election and choose itself. The rationale is that if the client can start an election, then it is alive.

\item \href{https://github.com/kenluck2001/DistributedSystemReseach/blob/master/textbook/two-phase-commit.c}{two-phase-commit.c}

The execution of the transactional functions happens at slightly different stages for the coordinator and the participants. The coordinator acts as the trigger and executes the "commit" action locally based on the votes, while the participants execute it upon the coordinator's command.

\textbf{Potential Issues}:
    \begin{itemize}
    \item Asymmetry in Commit Execution: While this model achieves the outcome of either all participating nodes (coordinator and participants) effectively committing or aborting, the timing and triggering of the actual transactional function execution are different. This might be a point of confusion or a simplification for the demonstration.
    \item No Confirmation from Participants: The coordinator sends COMMIT but doesn't wait for any acknowledgment from the participants that they have indeed committed. In a more robust system, you might have a third phase (though the name "two-phase commit" implies two phases) or some form of confirmation to ensure all participants have completed the action.
    \item Error Handling After Decision: If the coordinator successfully decides to commit and sends COMMIT messages, but one or more participants fail to execute the commit, the coordinator isn't aware of this failure in the current implementation.

    \end{itemize}

\item \href{https://github.com/kenluck2001/DistributedSystemReseach/blob/master/textbook/sequence-paxos4.c}{sequence-paxos4.c}

The C code implements the Sequence Paxos consensus algorithm using MPI to coordinate actions between processes acting as Clients, Proposers, Acceptors, and Learners. It's designed to agree on a sequence of values in a distributed system, maintaining order even with potential failures. The Client proposes a sequence of values, and the Proposers, Acceptors, and Learners work through the Paxos protocol for each value to achieve agreement.  Learners inform the Client of decided sequence positions, ensuring the Client proposes the next value in the sequence only after the previous one is confirmed.  This process continues until all values are agreed upon, demonstrating distributed consensus on an ordered sequence.

\item \href{https://github.com/kenluck2001/DistributedSystemReseach/blob/master/textbook/lamport1-majority-voting8.c}{lamport1-majority-voting8.c}

The C code simulates a distributed key-value store where processes use Lamport clocks and majority voting to handle concurrent updates.  MPI is used for communication, and each process maintains a local copy of the key-value store. When a process wants to update a value, it sends messages to other processes containing the key, the value, and its Lamport clock.  Processes update their local Lamport clocks based on received messages to track the order of events.  Each process gathers messages and uses the Lamport timestamps to determine the most recent (agreed-upon) value based on a majority, effectively resolving conflicts in a distributed setting.

\end{itemize}

\subsubsection{Defensive Programming}

One good practice is to wrap function calls to facilitate debugging and resilience of OpenMPI program.

\begin{verbatim}
#include <mpi.h>
#include <stdio.h>

#define SAFE_MPI_CALL(call) \
    do { \
        int ret = (call); \
        if (ret != MPI_SUCCESS) { \
            char error_string[MPI_MAX_ERROR_STRING]; \
            int length_of_error_string; \
            MPI_Error_string(ret, error_string, &length_of_error_string); \
            fprintf(stderr, "MPI Error (%s:%d): %s\n", __FILE__, __LINE__, error_string); \
            MPI_Abort(MPI_COMM_WORLD, ret); \
        } \
    } while (0)

int main(int argc, char** argv) {
    SAFE_MPI_CALL(MPI_Init(&argc, &argv));
    int rank;
    SAFE_MPI_CALL(MPI_Comm_rank(MPI_COMM_WORLD, &rank));
    printf("Hello from rank %d\n", rank);
    SAFE_MPI_CALL(MPI_Finalize());
    return 0;
}
\end{verbatim}

This code introduces a safety mechanism for MPI function calls through a macro named SAFE\_MPI\_CALL. When you wrap an MPI function call with this macro, it automatically checks if the function executed successfully. If an error occurred, the macro will fetch the specific error message from the MPI library, report it to the error output along with the exact location in your code where the failure happened, and then immediately stop the entire MPI program. This ensures that errors in MPI operations are clearly identified and prevent the program from continuing in a potentially unstable or incorrect state. Using this macro promotes more robust and easier-to-debug MPI applications by standardizing error checking. Unfortunately, there is an overhead introduced by the macros, but consistent error handling outweighs the potential downsides.

We have saved the best for the last. You must have noticed recurring themes in the book where we make abstractions. For example, our leader election uses a failure detector. Decomposing systems into components allows one to focus on a part of a problem at the time. People call it adding one more layer of indirection, it helps to reason about systems but avoid leaky abstractions at all costs. "Most problems in Computer science can be solved with another level of abstraction".


\section{Case Studies}

We describe our implementation of a barebone serverless framework for autoscaling in Subsection~\ref{autoscaling-casestudy} and an overview of distributed computing patterns in Subsection~\ref{distributed-computing-patterns}.

\subsection{Implementation of AutoScaling Framework (Serverless)}
\label{autoscaling-casestudy}

I've also developed a basic autoscaling framework that adds or removes compute nodes based on predefined CPU load thresholds. A drawback of this initial implementation is its non-incremental nature: we currently tear down existing nodes before adding new ones, which can lead to temporary memory spikes. Implementing autoscaling in OpenMPI requires careful engineering because the total number of MPI processes for communication groups must be set at the initialization. While OpenMPI groups are useful for limiting computation to specific processes (beneficial to avoid performance issues with too many communicating processes), this fixed-size requirement makes dynamic scaling difficult, and is why we've adopted our current node replacement strategy. See source code in \href{https://github.com/kenluck2001/DistributedSystemReseach/blob/master/textbook/autoscaling.py}{autoscaling.py}. The parameters for our implementations are arbitrarily chosen.

We detected CPU usage on each node using command \href{https://man7.org/linux/man-pages/man1/top.1.html}{top} run using \href{https://linux.die.net/man/3/execl}{execl} to run the \href{https://man7.org/linux/man-pages/man1/top.1.html}{top} executable available in the PATH usually and use popen to read the content from \href{https://man7.org/linux/man-pages/man1/top.1.html}{top} command. Replicate the CPU usage and estimate the average. Use as a heuristic to trigger process management.

Our implementation has Constants and helper functions.

~\textbf{Constants} \\*
\begin{itemize}
\item MIN\_THRESHOLD = 25: The lower CPU utilization threshold. If the average CPU load falls below this, the script will potentially reduce the number of MPI processes (though the current logic only increases).
\item MAX\_THRESHOLD = 75: The upper CPU utilization threshold. If the average CPU load exceeds this, the script will increase the number of MPI processes.
\item PULSE\_TIME = 300: The interval (in seconds) at which the script checks the CPU utilization and potentially scales the number of processes. This determines how frequently the autoscaling logic is evaluated (currently set to 5 minutes).
\end{itemize}

~\textbf{Helper Functions} \\*
\begin{itemize}
\item run\_command\_nonblocking(command, shell=False): Executes a given command (as a string or list) in the background without waiting for it to complete. It returns the subprocess. Popen object, which represents the running process. This allows the script to start a process and continue with other tasks.
\item get\_process\_output(process): Takes a subprocess.Popen object waits for the process to finish, and returns its standard output (stdout), standard error (stderr), and return code.
\item is\_process\_running(pid): Checks if a process with the given process ID (pid) is currently running on the system using psutil.pid\_exists().
\item kill\_process(process): Terminates a given process. It first tries a gentle terminate() and then forcefully kills it using kill() if it doesn't stop within a short time. It also prints messages indicating the process's status.
\item run\_process(process): Takes a subprocess.Popen object prints its PID and includes commented-out example code for potentially doing other work while the process runs or checking its status. It also includes a try...except...finally block to get the process output and ensure the process is killed in case of an error or completion.
\end{itemize}

~\textbf{autoscale\_processes() Function} \\*

    This is the core of the autoscaling logic.
\begin{itemize}
\item It initializes variables: command, process, pid to track the currently running MPI process, num\_of\_processes to the initial number of MPI processes (set to 8), and previous\_cpu\_avg to 0 to ensure the first iteration triggers an action.
\item It enters an infinite while True loop, which represents the continuous monitoring and scaling process.
\item Monitoring: Inside the loop, it first compiles (mpicc -g \href{https://github.com/kenluck2001/DistributedSystemReseach/blob/master/textbook/cpu-stats.c}{cpu-stats.c} -o cpu-stats) and then runs the cpu-stats MPI program with the current num\_of\_processes. It captures the standard output, which is assumed to be the average CPU load across the MPI processes.
    Decision Making: It calculates the absolute difference between the current and previous average CPU load. If this difference is greater than MIN\_THRESHOLD or if it's the first iteration (previous\_cpu\_avg == 0), it proceeds with scaling logic.
    
\item Scaling Logic:
        \begin{itemize}
        \item If the cpu\_avg is below MIN\_THRESHOLD (25\%), it checks if a process is currently running (pid and is\_process\_running(pid)) and kills it. It then sets num\_of\_processes to 8 and starts a new MPI process running sample.c with this reduced number of processes.
        \item If the cpu\_avg is between MIN\_THRESHOLD (25\%) and MAX\_THRESHOLD (75\%), it similarly kills any running process, sets num\_of\_processes to 16 and starts a new MPI process with this increased number of processes.
        \item If the cpu\_avg is above MAX\_THRESHOLD (75\%), it kills the running process, sets num\_of\_processes to 32, and starts a new MPI process with this further increased number of processes.
        \end{itemize}
\item Throttling: After the scaling decision (or if no scaling was needed), the script pauses for PULSE\_TIME (300 seconds) before checking the CPU load again.
\item Updating History: Finally, it updates previous\_cpu\_avg with the current cpu\_avg for the next iteration's change detection.
\end{itemize}

\subsubsection{Remarks} 

Another approach may involves group of processes. We upscale my union of groups and downscale by interesction similar to set theory. At the nodes still has to be initialized at the start, therefore, those unused nodes that were already spawn result in wasted resources.

For future improvements, depending on the application's requirements, it might be beneficial to avoid terminating existing processes when scaling up the number of resources. For instance, if we currently have 4 processes running and the CPU load increases, we could scale up to 8 processes to better manage the workload without interrupting the initial 4. The following code snippets illustrate how to independently launch two sets of 4 processes:

\begin{verbatim}
mpicc -g sample.c -o sample && mpiexec -n 4 ./sample
mpicc -g sample.c -o sample && mpiexec -n 4 ./sample
\end{verbatim}

However, when scaling down, we would need to explicitly terminate processes and maintain a record of the active process count to ensure accurate resource management.

\subsection{Distributed Computing Patterns}
\label{distributed-computing-patterns}


\subsubsection{Map-Reduce}

\begin{figure}[H] 
\centering
\includegraphics[scale=0.85]{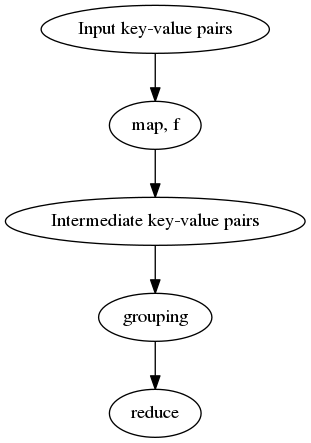}      
\caption{Map Reduce. }
\label{map_reduce}
\end{figure}

Key-value must be immutable. A map function accepts a key and outputs a value. The next operation is grouping (intermediate reduce operation) user-specific reduce (same value have aggregated value) map-reduce is a functional programming paradigm as shown in Figure~\ref{map_reduce}.

Spark key-value pair -resilient Distributed dataset (grouping)
lazy evaluation (caching -reducing computation grouping) e.g td-idf, page-rank
One possible way to approach it is to consider the problem in two cases
Case 1: rank 0 has all data and every other node is empty

\begin{itemize}
\item split the data in chunks
\item mpi-scatter to send to every process (or even mpi\_send). An advanced user may send to a group of users to handle intermediate nodes where secondary reduction may be performed on a subset of nodes.
\item Do computation between a subset of the node and set the result of the computation to a non\-overlapping set to aggregate the result.
\end{itemize}

Case 2: every node already.

\begin{itemize}
\item skip the mpi scatter phase in case 1.
\item Regroup appropriately.
\end{itemize}

The author encourages the user to implement map-reduce using the description provided in this section.

\subsubsection{Distributed Shared Primitive}
The shared variable is protected with a lock to protect a critical section and compare and swap can be used to update the critical section in a non-blocking manner.
I am not sure of how compare-and-swap works in mpi, my implementation of distributed shared primitive would be enhanced without blocking, which is more in line with our non-blocking approach in the core philosophy.
if you adjust the resilience of your algorithm to enhance fault tolerance within the limit provided by quorum, then remember to cancel every outstanding request to prevent livelocks.
All of our approaches to building a Distributed shared primitive are influenced by the work. One of our variations of a shared variable across processes made use of one-sided communication to pass messages between processes and always select the maximum. This works in the form of quasi-atomics and only works because our application makes use of the monotonic increasing property as an invariant. See an example of commented use in \href{https://github.com/kenluck2001/DistributedSystemReseach/blob/master/textbook/single-paxos3-snapshot.c}{single-paxos3-snapshot.c}.

\begin{verbatim}
int modify_var(struct mpi_counter_t *count, int valuein) {
    int *vals = (int *)malloc( count->size * sizeof(int) );
    int val;
    near_atomic_shared(count, valuein, vals);
    count->myval = MAX(count->myval, valuein);
    vals[count->rank] = count->myval;
    val = 0;
    for (int i=0; i<count->size; i++)
    {
        val = MAX(val, vals[i]);
    }
    free(vals);
    return val;
}
\end{verbatim}

Another variation is just using the latest value as the total value shared across every process. This logic makes it easier to implement a compare and swap in the future if we decide to go fully non-blocking. See an example of use in \href{https://github.com/kenluck2001/DistributedSystemReseach/blob/master/textbook/single-paxos3-snapshot.c}{single-paxos3-snapshot.c}.

\begin{verbatim}
int reset_var(struct mpi_counter_t *count, int valuein) {
    int *vals = (int *)malloc( count->size * sizeof(int) );
    int val;
    near_atomic_shared(count, valuein, vals);
    count->myval = valuein;
    vals[count->rank] = count->myval;
    val = count->myval;
    free(vals);
    return val;
}
\end{verbatim}

\subsubsection{Distributed Hashmap} 
We have used the Lamport clock to make a distributed hashmap where every node has a local map (key-value pair) and in agreement, every node has the same key value key based on quorum.

\section{Practical Considerations}
\begin{itemize}
\item Who can partake in a leadership election? It is possible to designate nodes with permissions levels. One may decide if only nodes with write permissions or read permissions can partake in the election.
\item Setting appropriate time drift to specific acceptable clock drift in your failure detectors.
\item A hybrid approach may include a centralized server to allow late-joining clients to participate in future elections and get past information.
\item Set range of rounds that a client can participate in an election
\item The number of clients that can participate in quorum using the limits of the synchronous, asynchronous, and partially Synchronous.
\item It is important to be mindful of failures and implement some form of bounded wait on the response and throw some timeout. It can also be useful for debugging failures. Livelock can be up in terminal conditions as a loop is not received to change to the terminal condition.
\end{itemize}

\subsection{Tips for Testing}
\begin{itemize}
\item Modularizing the source code and performing the unit test. Read more about the characteristics of concurrency bugs and inspire the tests to be written.
\item Identify invariants as relations and use metamorphic testing to test with higher coverage.
\item Add assert on bad conditions violations within the source code and scope within a condition flag that can be passed as a command-line argument to toggle the assertion when needed.
\item Experiment on a testbed using a cluster of VMs in a LAN managed with a DevOps tool such as Vagrant, see hints.
\end{itemize}

\subsection{Evaluation Metrics}
\begin{itemize}
\item Message complexity
    \begin{itemize}
    \item number of messages required to complete an operation
    \item  bit complexity (message length)
    \end{itemize}
\item Time complexity
    \begin{itemize}
    \item communication steps
    \item All communication steps take one unit. The time between send(m) and deliver(m) is at most one unit.
    \end{itemize}
\end{itemize}

\section{Exercises for the Readers}

\begin{enumerate}
\item Extend the implementation of sequence Paxos to work beyond 4 processes as is the limit of our demonstration in \href{https://github.com/kenluck2001/DistributedSystemReseach/blob/master/textbook/sequence-paxos4.c}{sequence-paxos4.c} using a shared distributed primitives. Take hints from \href{https://github.com/kenluck2001/DistributedSystemReseach/blob/master/textbook/single-paxos3-snapshot.c}{single-paxos3-snapshot.c} which work for any number of processes. The idea in our sequence Paxos implementation is to make the client send each item in an array one at a time, and once decided by the learner, send the index for the next item that the client has to send to the proposer.
\item Implement a three-phase commit as shown in SubSection~\ref{Atomic-Commit-Protocols}. Use the source code implemented for two-phase commit (\href{https://github.com/kenluck2001/DistributedSystemReseach/blob/master/textbook/two-phase-commit.c}{two-phase-commit.c}) as a guide.
\item Write tests for the Distributed algorithm discussed in the book. We will give the user the task of implementing tests as an exercise.
\item Implement telemetry for experimentation on characteristics of any algorithm discussed in this book. Hint: see our design of communication of telemetry in \href{https://github.com/kenluck2001/DistributedSystemReseach/blob/master/textbook/single-paxos3-snapshot.c}{single-paxos3-snapshot.c}.
\item Set up a testbed with a \href{https://github.com/mrahtz/mpi-vagrant}{simulated LAN with vagrant VM} for running mpi cluster.

\item Implement network shaping using VM to test out different performances in varying network bandwidths.
\item Improve the \href{https://github.com/kenluck2001/DistributedSystemReseach/blob/master/textbook/autoscaling.py}{autoscaling.py} using our OpenMPI framework. See our implementation in SubSection~\ref{autoscaling-casestudy}, then roll out yours and enjoy the challenge.
Hints: Use the ideas of incremental spawning of processes, rather than shutting every process down and restarting with the expected number of nodes.

\item This project involves training a fundamental linear regression model. The optimization method to be used is Stochastic Gradient Descent (SGD), a common algorithm detailed at \href{https://en.wikipedia.org/wiki/Stochastic_gradient_descent#Linear_regression}{Stochastic Gradient Descent}. The overarching goal is to create the initial building blocks for a distributed learning system, specifically focusing on Federated Learning. Federated Learning is a technique that allows multiple devices or organizations to train a machine learning model collaboratively without sharing their data.

    To simulate this distributed environment, we will use the OpenMPI framework. Within this framework, each running process is an independent 'node' or participant in the federated learning system.

    For the data used in training, there are two possible approaches:
    \begin{itemize}
    \item Decentralized Data Generation: Each individual node (MPI process) will generate its own set of randomized training data. This mimics a real-world federated setting where data is naturally distributed.
    \item Centralized Generation and Distribution: The primary node (the process with rank 0) will create the entire training dataset. Subsequently, this central node will distribute portions of the data to all other participating nodes via broadcasting.
    \end{itemize}

    This project emphasizes understanding the trade-offs of distributed learning, such as communication costs, data heterogeneity, and synchronization, with the primary goal of a basic, working federated linear regression implementation. The process involves each node training a local model on its data, followed by an aggregator collecting these pre-computed models to create a global model using model averaging (see \href{https://www2.stat.duke.edu/courses/Spring05/sta244/Handouts/press.pdf}{model averaging}).

    As a helpful starting point, consider the strategies for parallelizing linear algebra computations, as presented in Jeff Dean's bachelor thesis available at \href{https://drive.google.com/file/d/1I1fs4sczbCaACzA9XwxR3DiuXVtqmejL/view}{Jeff Dean's Bachelor thesis}. These techniques for distributing and combining computations can offer valuable insights~\footnote{
For potential future development and expansion of this project, consider exploring the following ideas: including \href{https://ai.facebook.com/blog/asynchronous-federated-learning/}{asynchronous late update} and \href{https://openreview.net/pdf?id=DiKT4rrUD9n}{dropout}}~\footnote{
\footnotesize \copyright Kenneth Emeka Odoh {\includegraphics[scale=0.06]{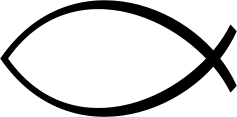}}. All contents of this book is covered under \href{https://kenluck2001.github.io/static/license.txt}{\includegraphics[scale=0.015]{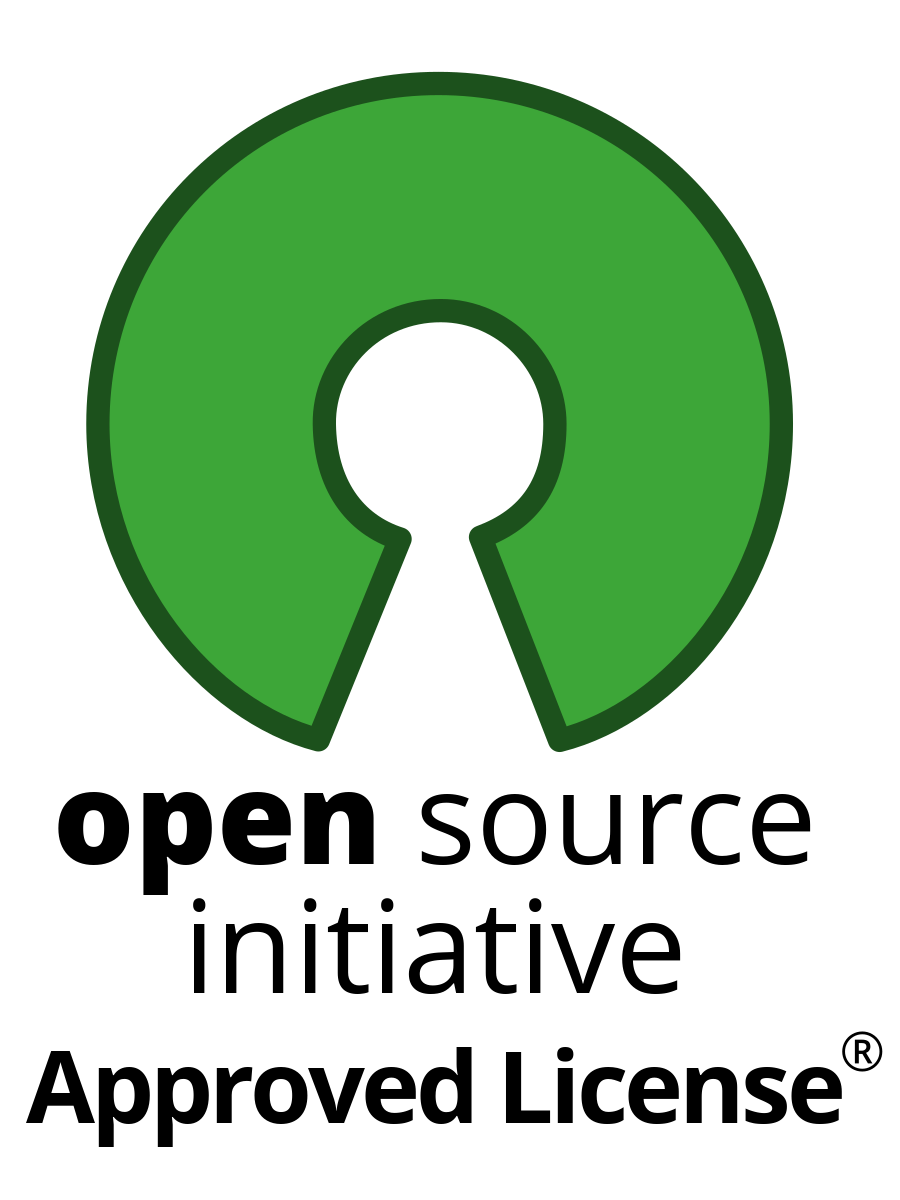}} \href{https://opensource.org/licenses/MIT}{license} \\Updated: \@submitdate}.

\end{enumerate}

\bibliography{report}
\bibliographystyle{plain}

\appendix

\end{document}